\documentclass[a4paper,11pt]{article}
\pdfoutput=1

\usepackage{jcappub}
\usepackage[T1]{fontenc}
\usepackage[utf8]{inputenc}
\usepackage{lmodern}

\usepackage{subfig}
\usepackage{epsfig}
\usepackage{graphicx}
\usepackage{color}
\usepackage{amssymb}
\usepackage{amsmath}
\usepackage{mathtools}
\usepackage{hyperref}
\usepackage{url}
\usepackage[normalem]{ulem}
\usepackage{afterpage}
\usepackage{float}
\usepackage{textcomp}

\definecolor{pdv}{rgb}{.7,0,.1}

\definecolor{valecol}{rgb}{0,0.5, 1.}

\definecolor{tcol}{rgb}{1.0,0.5, 0.5}

\arxivnumber{2409.09977} % Only if you have one
\title{\boldmath Redshift Drift fluctuations from $N$-body simulations}

\author[1]{{Pedro} Bessa}
\author[2,3,4]{Valerio Marra}
\author[3,4,5]{Tiago Castro}

\affiliation[1]{PPGCosmo, Universidade Federal do Espírito Santo, 29075-910, Vitória, ES, Brazil}
\affiliation[2]{Dep.\ de Física, Universidade Federal do Espírito Santo, 29075-910, Vitória, ES, Brazil}
\affiliation[3]{INAF -- Osservatorio Astronomico di Trieste, 34131 Trieste, Italy}
\affiliation[4]{IFPU -- Institute for Fundamental Physics of the Universe, 34151, Trieste, Italy}
\affiliation[5]{INFN – Sezione di Trieste, I-34100 Trieste, Italy}
\emailAdd{pedvbessa@gmail.com}

\abstract{Measurements of the redshift drift -- the real time variation of the redshift of distance sources -- are expected in the next couple of decades using next generation facilities such as the ANDES spectrograph at the ELT and the SKAO survey. The unprecedented precision of such observations will demand precise theoretical and numerical modeling of the effect in the standard $\Lambda$CDM cosmology. In this work, we use the \texttt{Gadget4} $N$-body code to simulate the redshift drift and its fluctuations in $\Lambda$CDM cosmologies, deriving the corresponding power spectra from a simulation with $1024^3$ particles in a $1\textrm{Gpc}\,h^{-1}$ box. Our results represent an initial step toward deriving the redshift drift fluctuation power spectra from $N$-body simulations and establishing a methodology for the statistical analysis of the redshift drift effect using data from future large-scale surveys. However, further work is required to refine the approach and achieve an accurate modeling of the redshift drift fluctuation power spectra.}


\begin{document}
\maketitle
\flushbottom

\section{Introduction}

 As cosmology develops further into its precision era, the theoretical and computational modeling of astrophysical and cosmological phenomena becomes ever more important in order to make proper sense of observations, break degeneracies, and probe new regimes. The capabilities of next- and current-generation surveys allow observational cosmology to probe astrophysical and cosmological phenomena previously thought unreachable and use these new probes to further test the standard cosmological model.
Among the forthcoming probes and facilities, the advent of state-of-the-art spectrographs such as ESPRESSO at the ELT~\cite{Martins_2022,cristiani_spectrographs_2023} and large radio surveys such as SKAO~\cite{ska1,ska2} will enable, within the next couple of decades, the first direct measurement of the time variation of the redshift of distant sources, known as the redshift drift. This measurement will provide model-independent constraints on the cosmological parameters of $\Lambda$CDM and its extensions in the local Universe through real-time cosmology, offering an important complement to distance-based probes. Real-time cosmology aims to infer cosmological parameters from the time variation of observables along the past light-cone of the observer~\cite{klockner2015realtimecosmology,Quercellini:2010zr}, and the redshift-drift detection would yield the first real-time determination of the local Hubble parameter.

The redshift drift offers a powerful new probe of cosmological dynamics, as it provides a model-independent test of the variation in the expansion rate of the Universe. Crucially, it does not rely on prior assumptions about the cosmic energy content, but instead depends solely on metric and kinematic properties. Its theoretical prediction was first derived  by Sandage and McVittie \cite{Sandage:1962,McVittie:1962} as a general consequence of non-static cosmological models. Its detection and measurement through the observation of the Lyman-$\alpha$ forest of distant quasars, was proposed by Loeb \cite{Loeb:1998bu}, in what is now known as the Sandage-Loeb test, although measurements from photometric surveys and strong gravitational lenses have also been investigated \cite{Kim:2014uha,Piattella_2017}.

With the possibility of measuring this effect in the near-future, and its use as a test of the cosmological model, research has focused on understanding the evolution and redshift dependency of the redshift drift in general cosmological space-times, in particular perturbed Robertson-Walker metrics \cite{heinesen_multipole_2020, Bessa_2023,Marcori:2018cwn,Kim:2014uha}; its use as a test of dark energy and alternative cosmological scenarios \cite{Heinesen_2021,Jain:2009bm,Melia:2024qha}; and as a probe of cosmic homogeneity \cite{koksbang_redshift_2022,Koksbang:2020zej,Koksbang:2021qqc,Koksbang:2019glb}. Important work has been done on forecasting the measurement of the effect through observations of the Ly$\alpha$ forest by facilities like the ELT \cite{Liske:2008ph,cooke_acceleration_2019,martins_cosmology_2023} and the SKAO \cite{marques_fundamental_2023,klockner2015realtimecosmology,bolejko_direct_2019}.

From the theoretical point of view, by extending the redshift drift definition to arbitrary spacetimes, Refs.~\cite{heinesen_multipole_2020,Heinesen_2021,Mishra:2013iu,Dunsby:2010ts} have allowed the calculation of perturbative and contaminating effects from cosmological inhomogeneities and local structures on the drift. In particular, Ref.~\cite{Bessa_2023} derived the gauge-invariant expression for the redshift drift in a perturbed Robertson-Walker spacetime and its power spectrum, which allows one to interpret future observations and its perturbative effects to first order, in full concordance with the standard cosmological model and its independent late and early time tests.

In scenarios where direct observational data is limited, simulations and numerical implementations provide a controlled framework to test hypotheses and explore the effects of various cosmological parameters on a given effect or observable. Modeling of the redshift drift and observational errors through $N$-body simulations was performed first in \cite{koksbang_redshift_2023}, and its power spectra was first numerically obtained through the use of Einstein-Boltzmann solvers in \cite{Bessa_2023}, and using relativistic numerical codes in \cite{Koksbang:2024xfr}.  There has been, however, a lack of direct modeling of the effect and its power spectra through $N$-body simulations in the $\Lambda$CDM setting, including peculiar velocity and acceleration effects. 

This work aims to fill this gap in the literature by using Newtonian $N$-body simulations and its mapping to relativistic cosmology as a nonlinear model for a Universe, and deriving the scale and redshift dependency of the drift through its power spectra obtained directly from simulations, providing a first estimate of the effect in the $\Lambda$CDM model from $N$-body simulations.

The paper layout is as following:
In Section~\ref{sec:dz}, we review the mathematical background needed to derive the redshift drift fluctuations in a Robertson-Walker cosmology. In Sections~\ref{sec:pdz} and~\ref{sec:sims}, we review the derivation of the redshift drift fluctuations power spectrum in $\Lambda$CDM and its approximations, as well as the map between Newtonian perturbations obtained from the $N$-body simulations and the theoretical relativistic quantities. In Section~\ref{sec:setup}, we detail our simulation parameters and set-up and the numerical derivation of the power spectra. We also detail our error estimation and modeling. In Section~\ref{sec:conc}, we write our conclusions and discuss the results.

%Central to our methodology is the derivation of redshift drift fluctuation power spectra from these simulations. This is achieved by integrating existing theoretical models with our computational simulations. The use of the flat sky approximation enables the extraction of spatial power spectra from simulation volumes, providing insight into the distribution and evolution of cosmic structures at large scales. Concurrently, the Limber approximation is applied to derive velocity and acceleration power spectra, essential for understanding the kinematic aspects of redshift drift.

%The study then focuses on comparing the outcomes of these simulations with the predictions made by Einstein-Boltzmann solvers. This comparison is crucial for assessing the accuracy of our simulation models and the validity of theoretical predictions concerning redshift drift. By aligning our findings with established theoretical models, we aim to either confirm or challenge current understanding of cosmic expansion dynamics.

\section{Redshift drift and fluctuations}
\label{sec:dz}

 We model the redshift drift fluctuations using linear cosmological perturbation theory on a homogeneous and isotropic background, given by Robertson-Walker metric. Since we are interested in the concordance $\Lambda$CDM model, we focus on the flat case. First, we obtain its expression in flat Robertson-Walker and then generalize it to arbitrary spacetimes, such that the fluctuations become a particular case where the metric is given by the perturbed Robertson-Walker metric.
In a flat Robertson-Walker spacetime, the line element is given by
\begin{equation}
    \label{eq:rw_flat}
    ds^2 = -a^2(t)[dt^2 + \delta_{ij}dx^i dx^j ],
\end{equation}
where we denote cosmic time by $\tau$ and conformal time by $t$.  The expression for the redshift drift can be easily derived through the cosmic time derivative of the redshift \cite{Loeb:1998bu}
\begin{equation}
    \label{eq:dz_back}
    \frac{dz}{d\tau_o} = \frac{d}{d\tau_o}\left(\frac{a_o}{a_s}\right) =\frac{1}{d\tau_o}\left[\frac{a_o(1+H_od\tau_o)}{a_s(1+H_s (a_s/a_o)d\tau_o)}-\frac{a_o}{a_s}\right] = H_o(1+z)-H_s \,.
\end{equation}
where we use the subscripts $o$ and $s$ to denote measurements at the observer and source, respectively.
The expression for the drift above shows that observing the redshift of a source at different observer times $\tau$ provides a measure of the Universe’s properties at the corresponding redshift. From Eq.~\eqref{eq:dz_back}, one can show that the redshift drift is positive when the Universe is undergoing accelerated expansion, as in the late-time $\Lambda$CDM phase \cite{Martins:2016bbi}. At the background level, this is true for any given observer since we have spacetime homogeneity and isotropy, but once perturbations are introduced, the existence of cosmic structures changes the path of light rays through their gravitational potential and peculiar velocities, such that Eq.~\eqref{eq:dz_back} has to be modified.

In \cite{heinesen_multipole_2020}, the author derived a generalization of expression \eqref{eq:dz_back} for the redshift drift in an arbitrary spacetime, relying only on the assumption of a $3+1$ slicing of the spacetime. For an observer $o$ with 4-velocity $u^\mu$ and a light ray emitted from source $S$ with tangent 4-vector $k^\mu$, the redshift drift measured by the observer is 
\begin{align}
    \frac{dz}{d\tau_o} &= (1+z)\left[\mathfrak{h} -\frac{\nabla_e E}{E}\right]_o-\left[\mathfrak{h}-\frac{\nabla_e E}{E}\right]_s\,,~~\text{ with} \nonumber\\
    \mathfrak{h}&:= \frac{1}{3}\theta+\sigma_{\mu\nu}e^\mu e^\nu -a_\mu e^\mu\,, \label{eq:generalised_dz}
\end{align}
where we have the usual kinematic quantities -- the expansion, acceleration, and traceless expansion tensor -- along the observer's congruence, defined respectively by
\begin{equation}
    \theta = \nabla_u u, \quad a_\mu = \nabla_\mu\theta, \quad \sigma_{\mu\nu} = \nabla_{(\mu}u_{\nu)}-\frac{1}{3}\theta h_{\mu\nu},
\end{equation}
where $h_{\mu\nu} = g_{\mu\nu}+u_\mu u_\nu$. Expression~\eqref{eq:generalised_dz}, in the case of the metric~\eqref{eq:rw_flat} simplifies to~\eqref{eq:dz_back}. We will use it to derive the expression for the drift in a Universe with perturbations.

\subsection{Redshift drift in perturbed Robertson-Walker}

We now consider the case of a Universe with inhomogeneities in the matter distribution and metric fields, treated as small perturbations around the background metric and described within the framework of a perturbed Robertson–Walker spacetime. In the longitudinal gauge, the line element for this metric is given by \cite{durrer_2020}
\begin{equation}
\label{eq:rw_perturb}
ds^2 = -a^2(t)^2(1+2\Psi)dt^2 + a^2(t)(1-2\Phi)\delta_{ij}dx^i dx^j,
\end{equation}
where the potentials $\Phi$ and $\Psi$ are the gauge-invariant Bardeen potentials. We also write the observer's 4-velocity in this spacetime as
\begin{equation}
\label{eq:obs_vel}
u^\mu = \frac{1}{a}(1-\Psi,v^i), \quad v_i \equiv \partial_i v,
\end{equation}
with $v$ the velocity potential, and $u^\mu$ a global defined congruence of comoving observers.

Following fully gauge-invariant cosmological perturbation theory,  Ref.~\cite{Bessa_2023} obtained the gauge-invariant expression for the redshift drift \eqref{eq:generalised_dz} in the perturbed Robertson-Walker metric, which here we express in terms of the longitudinal gauge used in \eqref{eq:rw_perturb}
\begin{eqnarray}
\label{eq:dz_longitudinal}
    \frac{dz}{d\tau_o} &=& (1+z)\Bigg[-\mathcal{H}+\Dot{\Psi}+\bar{n}^i\partial_i(\Dot{v}+\mathcal{H}v) +\frac{\Dot{\mathcal{H}}}{\mathcal{H}}(\Phi-\bar{n}^i\partial_iv)+
	 \nonumber \\
  && \qquad\qquad \left(\frac{\Dot{\mathcal{H}}-\mathcal{H}^2}{\mathcal{H}}+\bar{n}^i\partial_i \right)\int_{s}^{o}\hspace{-1.3mm}d\lambda\left(\Dot{\Phi}+\Dot{\Psi}\right) \Bigg]\,,
\end{eqnarray}
where $-\bar{n}^i$ is the space-like vector of the source direction. 

In deriving this expression, we neglected observer-dependent dipole terms, namely the observer's peculiar velocity and gravitational potential, as we are focused in survey and simulations. Furthermore our main interest is on the source's peculiar motion, as the perturbative uncertainties and fluctuations contributing to the redshift drift of sources are mainly related to their peculiar velocities and local environment. The observer-dependent terms are important in and of themselves and have been previously measured in \cite{Inoue:2019qvy} and its impact on redshift drift measurements have also been discussed therein. In a realistic survey, such terms can be constrained and mitigated by correlation with local motion measurements such as described in \cite{Inoue:2019qvy}, since the effects are of the order of the cosmological ones. The monopole terms are unmeasurable at the observer as per \cite{durrer_2020}.
This observer dependent drift and its impact on measurements of the redshift drift has been studied in \cite{Inoue:2019qvy}.

The potential terms in expression \eqref{eq:dz_longitudinal} have been shown to be orders of magnitude smaller than the spatial dependent and velocity terms, particularly in the redshift range $z\in [0,1.5]$ in which we expect to measure the drift effect \cite{Bessa_2023}, and thus we assume that they are subdominant to the other perturbative quantities and do not contribute to the overall fluctuations of the drift. Focusing on terms containing the velocity and its time derivatives, and by defining $\frac{d\bar z}{d\bar\tau_o}$ as the background value of the redshift drift given by \eqref{eq:dz_back}, we obtain the expression for the perturbations of the redshift drift 
\begin{align}
\label{eq:dz_fluct}
    \delta\dot{z}&:=\frac{\frac{dz}{d\tau_o}-\frac{d\bar z}{d\bar\tau_o}}{\frac{d\bar z}{d\bar\tau_o}} \nonumber\\
    &\approx -\frac{1}{\mathcal{H}}\Bigg[\bar{n}^i \Dot{v}_i
    -\left(\frac{\Dot{\mathcal{H}}-\mathcal{H}^2}{\mathcal{H}}\right)\bar{n}^i v_i\Bigg]\;,
\end{align}
where we note again that all observer and potential terms were neglected.
In the following sections, we will derive the power spectrum of the signal \eqref{eq:dz_fluct} as a function of redshift measured by the observer.

\section{Redshift drift power spectrum}
\label{sec:pdz}

The redshift drift fluctuations power spectrum can be obtained directly from the primordial metric fluctuations power spectrum through the potential and velocity transfer functions and Einstein's equations, and this has been implemented and checked numerically in \cite{Bessa_2023} through Einstein-Boltzmann solvers. Here we briefly describe how one can obtain the redshift-drift power spectra from the primordial perturbations.

The primordial spectrum of curvature perturbations $P_\Psi$ generated by primordial fluctuations formed through a standard inflationary mechanism in the early Universe can be written as \cite{durrer_2020}
\begin{equation}
\label{eq:prim_spec}
k^3P_\Psi \equiv 2\pi^2\left(\frac{3}{5}\right)^2\mathcal{P_R} =(2\pi)^3\delta(\bold{k}-\bold{k}')A\left(\frac{k}{k_*}\right)^{n_S-1}, 
\end{equation}
where $k_*$ is a pivot scale that characterizes the perturbation scale, $A$ is the primordial amplitude of the Bardeen Potential $\Psi$ fluctuations, and $n_S$ is the spectral index defining the scale dependency of the perturbations. $\mathcal{P_R}$ is the scale-invariant primordial spectrum, which is related to the Fourier transform of the 2-point correlation function of the primordial fluctuations by $2\pi^2\left(\frac{3}{5}\right)^2\mathcal{P_R} = k^3\langle\Psi_\text{in}(k)\Psi^*_\text{in}(k')\rangle$.
The measured primordial amplitude, obtained via observation of the CMB is the primordial curvature fluctuation amplitude $A_S$, related to $A$ via \cite{durrer_2020}
\begin{equation}
\label{eq:prim_amp}
A = 2\pi^2\left(\frac{3}{5}\right)^2 A_S.
\end{equation}

The cosmological parameters $A_S$, $n_S$ have been measured and constrained through observations of the CMB by the Planck satellite, and an up-to-date review of their values can be found in \cite{Planck:2018vyg}. The values for these parameters used throughout this paper can be found in Table~\ref{tab:parameters}.

The value of the Bardeen potential at a given redshift $\Psi(z)$ is related to the primordial fluctuations $\Psi_\text{in}(k)$ through the Transfer Function $T_\Psi(k,z)$ and the growth factor $D_+(z)$, which measures cosmological structure formation up to redshift $z$, via \cite{Dodelson:2003ft}
\begin{align}
\label{eq:transfer_1}
\Psi(k,z) = T_\Psi(k,z)\Psi_\text{ini}(k) = \frac{9}{10}\frac{D_+(a)}{a}T(k)\Psi_\text{ini}(k).
\end{align}
The Transfer Function $T(k)$ is characterized by the evolution of the cosmological model and its parameters and can be obtained either analytically through an ansatz \cite{Eisenstein_1999}, or numerically using Einstein-Boltzmann solvers. Einstein's Equations allows one to use the Transfer Function $T(k)$ to obtain the fields $v$, $\Phi$ and $\Psi$ at a given redshift from the primordial power spectrum $\mathcal{P_R}$ using the relations \cite{Bonvin_2011}
\begin{subequations}
\begin{align}
    V(k,z)&= T_V(k,z)\Psi_\text{ini}(k),\\
    \Phi(k,z)&= T_\Phi(k,z)\Psi_\text{ini}(k),\\
    T_\Phi&=  T_\Psi,\label{eq:T_phi}\\
    T_V&= \frac{2a}{3\Omega_m}\frac{k}{\mathcal{H}_0^2}\left(\mathcal{H}T_\Psi +\Dot{T}_\Psi   \right),\label{eq:T_v}\\
    kT_\Psi&=\dot T_V +\mathcal{H} T_V \label{pec_acc}.
\end{align}
\end{subequations}

Applying these relations to \eqref{eq:dz_longitudinal} and decomposing observed fields in spherical harmonics $Y_{\ell m}(\bold{n})$ over the celestial sphere with unit vector $\bold{n}$, the observed angular power spectrum of the redshift drift fluctuations at redshift $z_S$ is given by
\begin{equation}
\label{eq:cl_theo}
    C^{\delta \dot{z}}_\ell(z_s) = \frac{2A}{\pi}\int \frac{dk}{k}\left(\frac{k}{k_*}\right)^{n_S-1}|F_\ell(k,z_s)|^2,
\end{equation}
where the kernel $F_\ell(k,z_S)$ is given by
\begin{equation}
    F_\ell(k,z_s) \approx -\frac{1}{\mathcal{H}}\Bigg[j_\ell(kr_s)\left( \dot{T}_\psi+\frac{\dot{\mathcal{H}}}{\mathcal{H}}T_\Psi\right)+ j'_\ell(kr_s)\left( \dot{T}_V-\frac{\Dot{\mathcal{H}}-\mathcal{H}^2}{\mathcal{H}}T_V\right)\nonumber\Bigg],
    \label{eq:F_ell}
\end{equation}
where $j_\ell$ are the spherical Bessel functions of order $\ell$, and $r_S$ is the comoving radial distance from the observer to the source. Here, we neglect the time-varying integral terms, which are small, while keeping the potential terms.

\section{Power spectrum from $N$-body simulations}
\label{sec:sims}

To test the theoretical predictions for the redshift drift power spectra $C^{\delta \dot{z}}_\ell(z_s)$ derived in the previous section, we employ the \texttt{Gadget4}\footnote{\url{https://wwwmpa.mpa-garching.mpg.de/gadget4/}} cosmological code \cite{Springel:2020plp} to conduct a Newtonian $N$-body simulation. Additionally, we use a modified version of the \texttt{CLASS} Einstein-Boltzmann solver \cite{DiDio:2013bqa,CLASS2}, previously introduced in \cite{Bessa_2023}, to validate our results and provide numerical computations for the power spectra.

In the longitudinal gauge the metric potentials in \eqref{eq:rw_perturb} and the velocities in \eqref{eq:obs_vel} map directly to the Newtonian gravitational potential $\psi$ and the particles' peculiar velocity vector $v_i$ \cite{Green:2011wc}, allowing us to treat them as the particle fields in a Newtonian $N$-body simulation. The theoretical prediction for the perturbed redshift drift field \eqref{eq:dz_longitudinal}, obtained using first-order cosmological perturbation theory, can then be directly mapped to the fields from the simulation  with negligible correction terms up to the Hubble scale, as first shown in \cite{Green:2011wc} to linear order.  To obtain the power spectra at different redshifts from the \texttt{Gadget4} simulations, we then follow this correspondence, formalized in the dictionary  between relativistic and Newtonian cosmologies first defined in section~II.C of~\cite{Green:2011wc}.

The angular power spectrum at redshift $z$ of an arbitrary 3D field $X$ is related to its power spectrum through the expressions \cite{Dodelson:2003ft}:
\begin{align}
    C^X_\ell(z) &=\frac{2}{\pi} \int_0^\infty dk k^2P^X_\text{obs}(k,z)W_{r,\ell}^2 (k) = 4\pi\int_0^\infty \frac{d k}{k}\Delta^2(k,z) W_{r,\ell}^2( k ),  \label{eq:cl_obs} \\
    W_{r,\ell} (k) &= \int_0^r d\bar{r} W(\bar r)   j^{(i)}_\ell ( k \bar r)\label{eq:window_func},
\end{align}
where $W(\Bar{r})$ is an arbitrary window function used to filter a certain range of redshifts or scales, and $j^{(i)}_\ell$ is the $i$-th derivative of the spherical Bessel function $j_\ell$, which appears in the Fourier transform of direction-dependent fields $X(\bold{n})$.
The dimensional power spectrum is related to the dimensionless one by $P^X_\text{obs}(k,z) = \frac{2\pi^2}{k^3} \Delta^2(k,z)$.

The two-point correlation function, decomposed into spherical harmonics, of the redshift drift fluctuation field $\delta\dot{z}$, when neglecting potential terms that are at least an order of magnitude smaller, is given by
\begin{align}
    \label{eq:drift_corr}
    \langle\delta\dot{z}(n,z)\delta\dot{z}(n',z) \rangle =& \sum_{\ell} \frac{2\ell+1}{4\pi}P_\ell(\bold{n}\cdot\bold{n}')\times \nonumber\\
    &\frac{2}{\pi}\int\frac{dk}{k}k^3\times k^2\left[\frac{1}{\mathcal{H}^2}P_{\dot{v}} -2\left(\frac{\Dot{\mathcal{H}}-\mathcal{H}^2}{\mathcal{H}^3}\right)P_{\dot{v}v} + \frac{\left(\Dot{\mathcal{H}}-\mathcal{H}^2\right)^2}{\mathcal{H}^3}P_{v}\right]j'\left(kr(z) \right)\nonumber\\
    =& \sum_{\ell} \frac{2\ell+1}{4\pi}P_\ell(\bold{n}\cdot\bold{n}')\times C_\ell^{\delta\dot{z}}(z),
\end{align}
which allows us to write the angular power spectrum for the redshift drift fluctuations in the form of equation \eqref{eq:cl_obs}:
\begin{align}
    \label{eq:cl_drift}
    &C_\ell^{\delta\dot{z}}(z) = \frac{2}{\pi} \int_0^\infty dk k^2P^{\delta\dot{z}}_\text{obs}(k,z)W_{r,\ell}^2 (k),\,\quad W_{r,\ell} (k) = \int_0^{r(z)} d\bar{r} W(\bar r)   j'_\ell ( k \bar r)\notag\\
    &P^{\delta\dot{z}}_\text{obs}(k,z) = k^2\left[\frac{1}{\mathcal{H}^2}P_{\dot{v}} -2\left(\frac{\Dot{\mathcal{H}}-\mathcal{H}^2}{\mathcal{H}^3}\right)P_{\dot{v}v} + \frac{\left(\Dot{\mathcal{H}}-\mathcal{H}^2\right)^2}{\mathcal{H}^3}P_{v}\right],
\end{align}
where we used implicitly the relations in Fourier space between the velocity potential and vector $V(k,z) = -ikv(k,z)\implies -ikV(k,z) = k^2 v(k,z)$.

\subsection{Limber Approximation}
For scales $\ell\gg 1 $, it is usual to adopt the Limber approximation, which we detail in appendix~\ref{sec:limber}. Using this approximation, \eqref{eq:cl_obs} can be further simplified to 
\begin{equation}
\label{eq:cl_limber}
    C^{X}_\ell(x) = \frac{2}{\pi} \int_0^\infty \frac{d k}{k}\Delta^2(k,z) W_{r,\ell}^2( k ) \approx \frac{1}{(\ell+1/2)^3}\int d\bar{r}\bar{r}\Delta^2 \left (\frac{\ell+1/2}{\bar{r}},z \right)W^2(\bar r).
\end{equation}
For distance modes satisfying $\ell \gtrsim \ell_c = \bar{r}/\Delta \bar{r}$ the Limber approximation has been shown to be accurate to smaller than $1\%$ accuracy levels in comparison to the fully integrated power spectra \cite{Dodelson:2003ft,durrer_2020}. We shall make use of this approximation as a consistency check when validating the simulations.

Furthermore, we use a top-hat window function, which filters modes outside a comoving size $\Delta \bar{r}$. Modes $k$ such that $\ell\gtrsim \frac{k\Delta\bar r}{2\pi\bar r}$ are integrated out of the angular power spectra. The explicit form of this window function is given by
\begin{align} \label{sel-fun}
W_{r}( \bar r)= \left\{\begin{array}{ll}
1/\Delta r & \; \mbox{ for } \;   r_1  < \bar r < r_2,\qquad \Delta r = r_2-r_1 \\
0 & \;  \mbox{ otherwise }.
\end{array}\right. \!\!\!\!
\end{align}
One can also interpret the window function above as selecting fluctuations inside the redshift slice $\Delta z \approx H(z)\Delta r/c$ centered on the redshift $z$.

This window function, when inserted in equation \eqref{eq:cl_limber}, allows one to simplify further the observed angular power spectrum expression in the Limber approximation. By selecting a small distance slice $\Delta \bar{r}$ of the observed cosmic region, we have
\begin{equation}
\label{eq:cl_approx}
    C_\ell^{X}(z) \approx \frac{1}{(\ell+1/2)^3}\int d\bar{r}\bar{r}\Delta^2 \left(\frac{\ell+1/2}{\bar{r}},z \right)W^2(\bar r) \xrightarrow{\Delta \bar r/\bar r\ll 1} \frac{\ell_c}{(\ell+1/2)^3}\Delta^2 \left (\frac{\ell+1/2}{\bar{r}},z \right).
\end{equation}
Equation \eqref{eq:cl_approx} allows simple implementation of the theoretical power spectrum \eqref{eq:cl_obs}, and although we use the \texttt{FFTlog-and-beyond}\footnote{\href{https://github.com/xfangcosmo/FFTLog-and-beyond}{github.com/xfangcosmo/FFTLog-and-beyond}} \citep[][]{Fang:2019xat} routine to calculate the integrals \cite{Fang_2020}, at smaller scales the Limber approximation provides consistency check for the numerical power spectra.

We note that approximations \eqref{eq:cl_limber} and \eqref{eq:cl_approx} are valid for integrals containing no derivatives of the Bessel functions, such as for the potential terms in \eqref{eq:cl_theo}. In the final expression \eqref{eq:cl_drift} the Limber approximation is not a good approximation due to the different oscillating behavior of the spherical Bessel function derivatives.

\begin{table}[h]
\begin{center}
\begin{tabular}{|l|l|l|}
\hline
Parameter& Value & Units\\
\hline
$N_\text{particles} $&$1024^3$& - \\
$m_\text{part}$ & $7.8\cdot 10^{10}$ & $M_\odot /h$ \\
$\text{Box Size} $& 1000 & Mpc/$h$\\
$\text{Grid Size} $&$2048^3$& - \\
$\epsilon_\text{soft}$ & 0.024 & Mpc/$h$\\
$\Omega_b$&$0.022383$&$h^{-2}$\\
$\Omega_{\text{CDM}}$&$0.12011$&$h^{-2}$\\
$h$&$0.687$&$10^{-2}{\rm\, km/s/Mpc}$\\
%$A_S$&$2.100549\cdot10^{-9}$& - \\
$n_S$&$1$& - \\
$z_{\text{eq}}$&$3387$& - \\
$k_{\text{eq}}$&$0.010339$& $h/{\rm Mpc}$\\
$\sigma_8$&$0.8$& -\\
$\Delta z$&$0.1$ & -\\
\hline
\end{tabular}
\end{center}
\caption{Parameters used to generate the $N$-body simulation with \texttt{Gadget4} and numerical spectra through Einstein-Boltzmann solvers.}
\label{tab:parameters}
\end{table}

\section{Simulation setup and derived power spectrum}
\label{sec:setup}

 To model the observed power spectra of the redshift drift fluctuations \eqref{eq:cl_drift} we perform a Newtonian $N$-body simulation using the \texttt{Gadget4} code. Our simulation was run on a $1\,{\rm Gpc}\,h^{-1}$ Box with $1024^3$ particles, and initial conditions were generated via the built-in {\texttt{NGEN-ic}} code at redshift $z=49$.
 Cosmological parameters, particle mass, and gravitational softening $\epsilon_\text{soft}$ can be found in Table~\ref{tab:parameters}. We produced three snapshots at the redshifts $z= 0.5,\,1,\,2$. For the window function, we use a redshift slice of $\Delta z = 0.1$ and a top-hat window function, which corresponds to a comoving size of $0.1c/H(z)$~Mpc for the projected field in equation \eqref{eq:cl_obs}.

\subsection{Validation}
\label{sec:validation}

\begin{figure}
    \centering
    \includegraphics[width=0.5\linewidth]{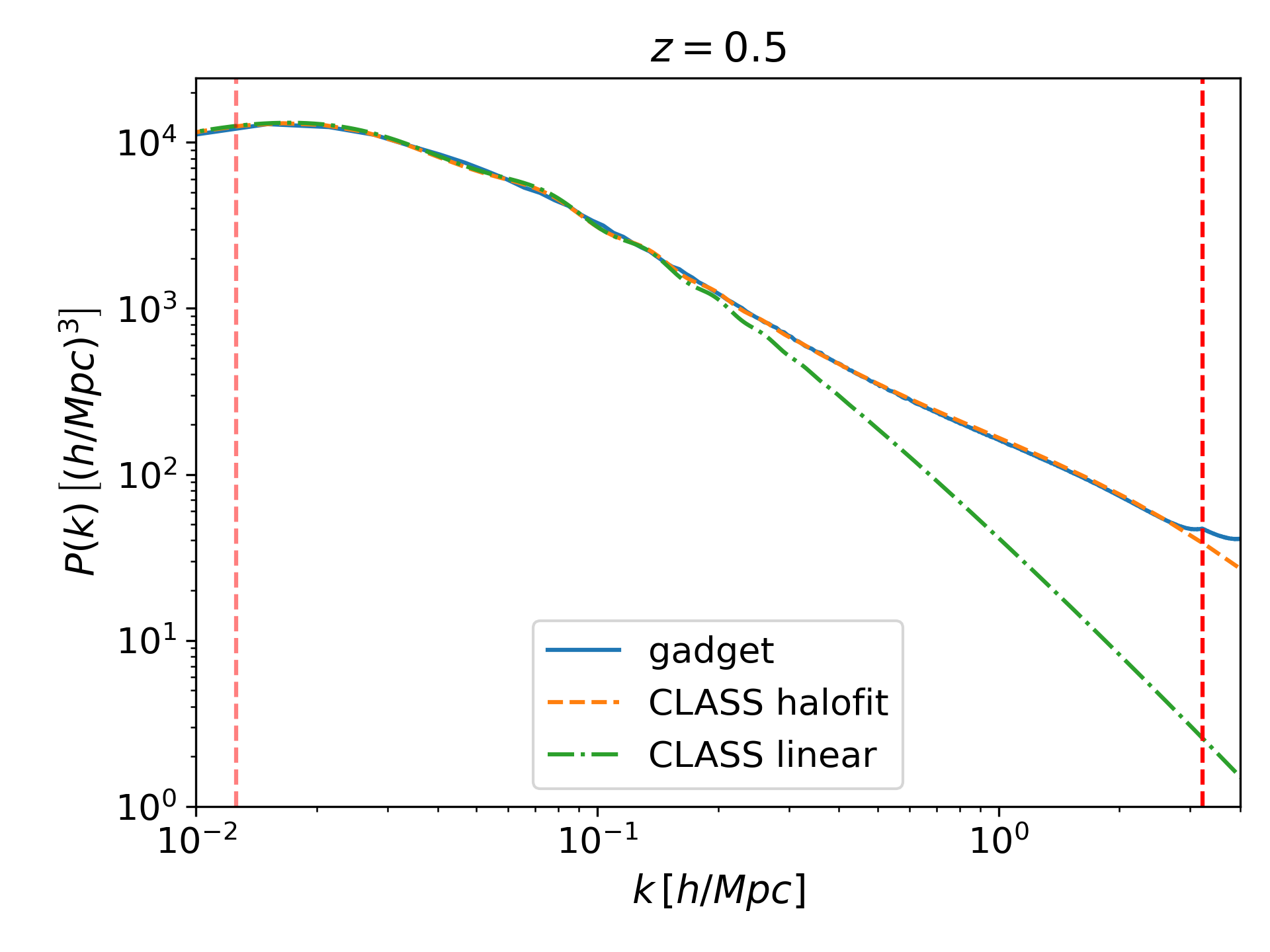}\hfil
    \includegraphics[width=0.5\linewidth]{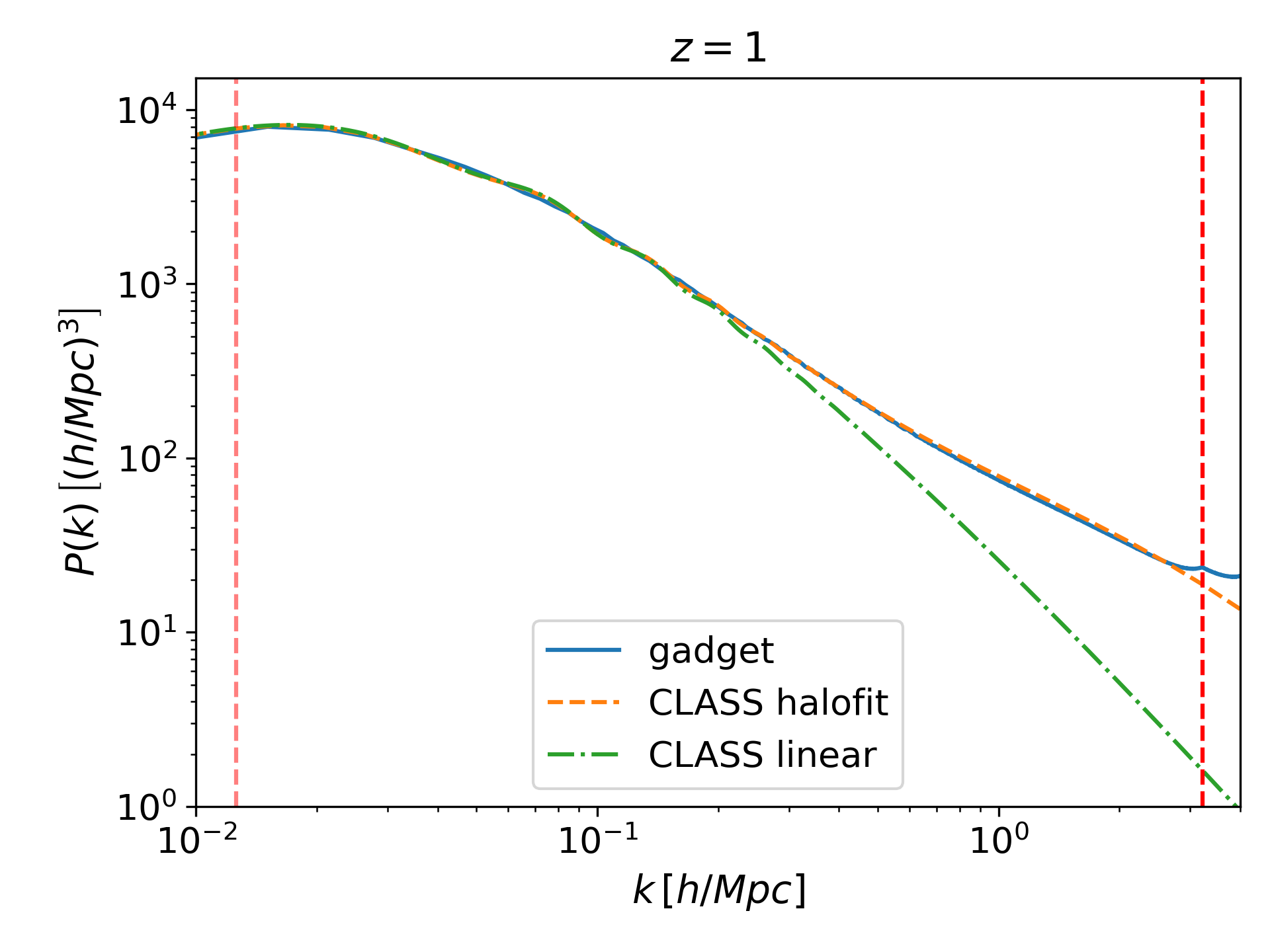}
    \includegraphics[width=0.5\linewidth]{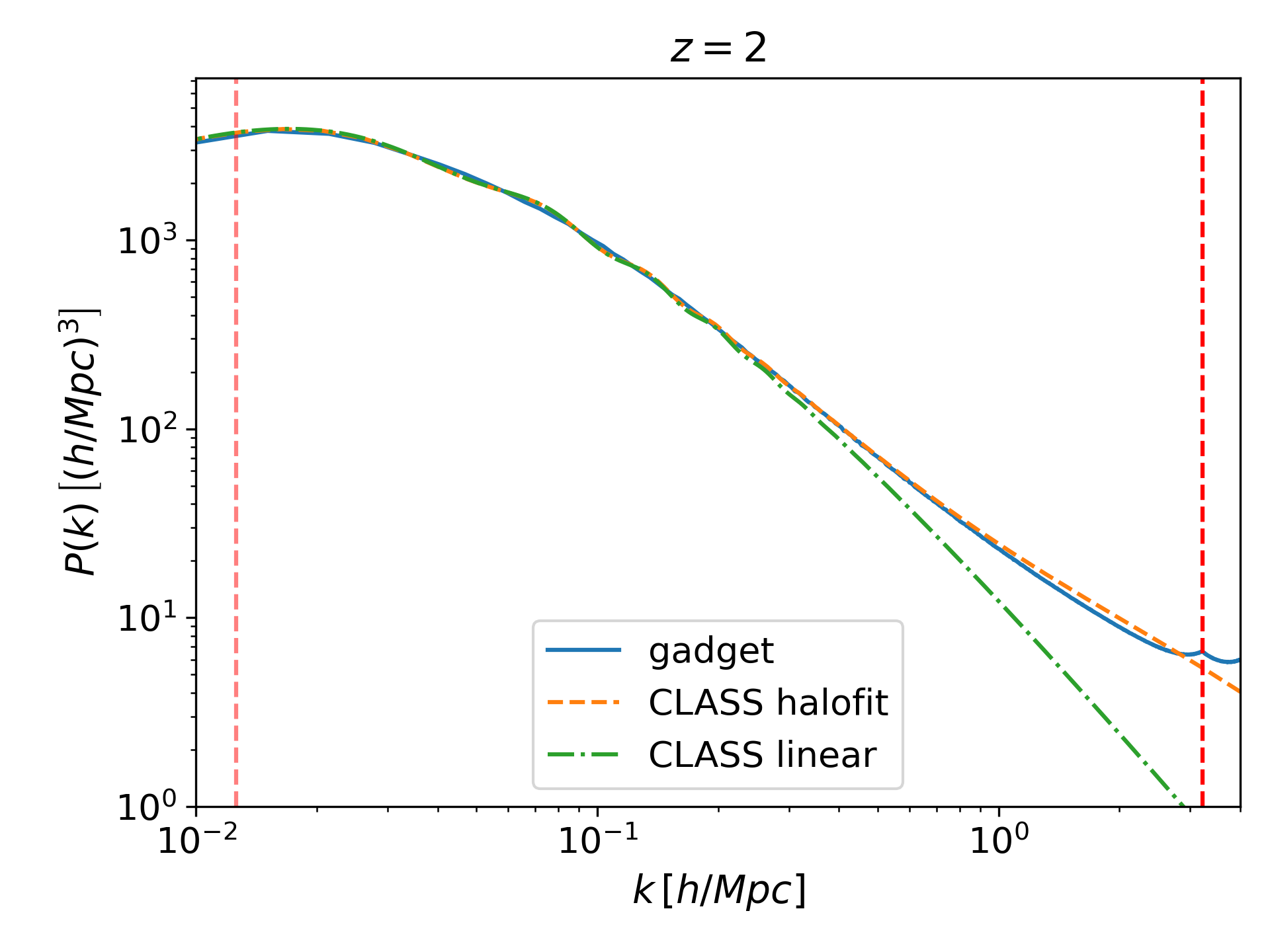}
    \caption{Matter power spectra for redshifts $z=0.5\,,1,\,2$. In the figures we compare two routines against the power spectra obtained from the \texttt{Gadget4} simulations: the halofit routine and  the linear matter power spectrum from \texttt{CLASS}.
    The vertical red lines mark $2k_\text{fund}$ and $k_\text{Nyq}/2$.}
    \label{fig:ps_validation}
\end{figure}

\begin{figure}
    \centering
    \includegraphics[width=0.5\linewidth]{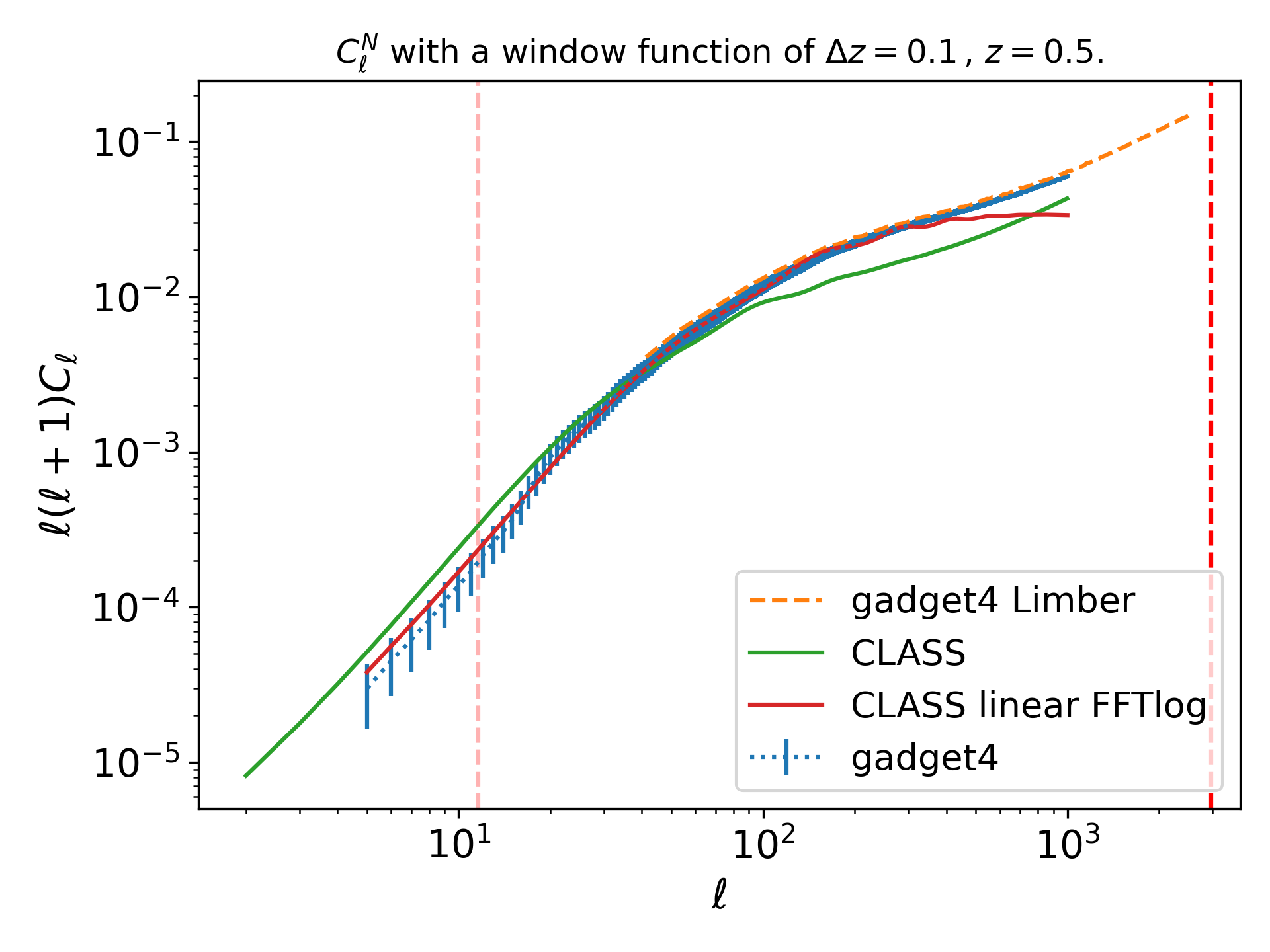}\hfil
    \includegraphics[width=0.5\linewidth]{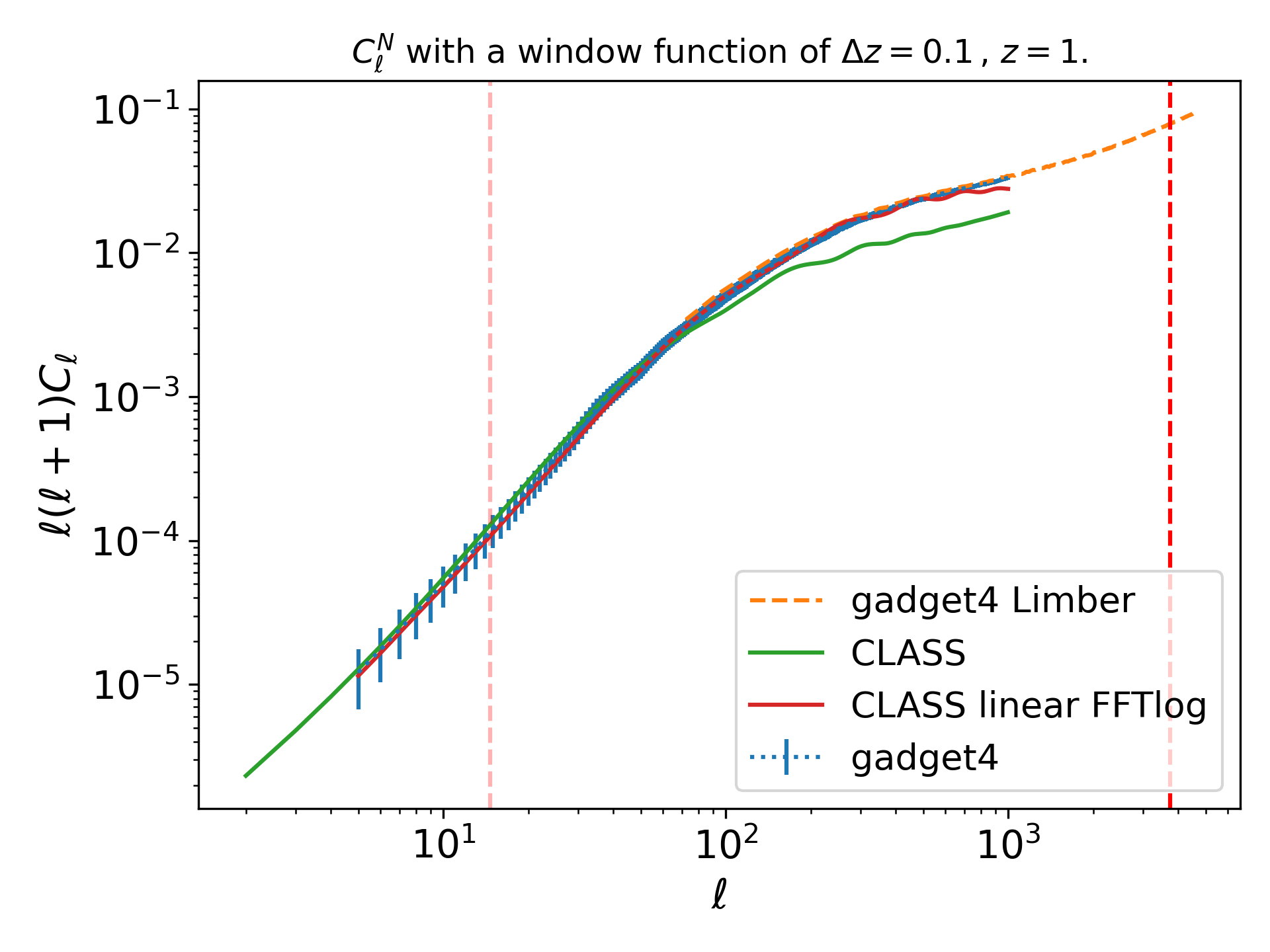}
    \includegraphics[width=0.5\linewidth]{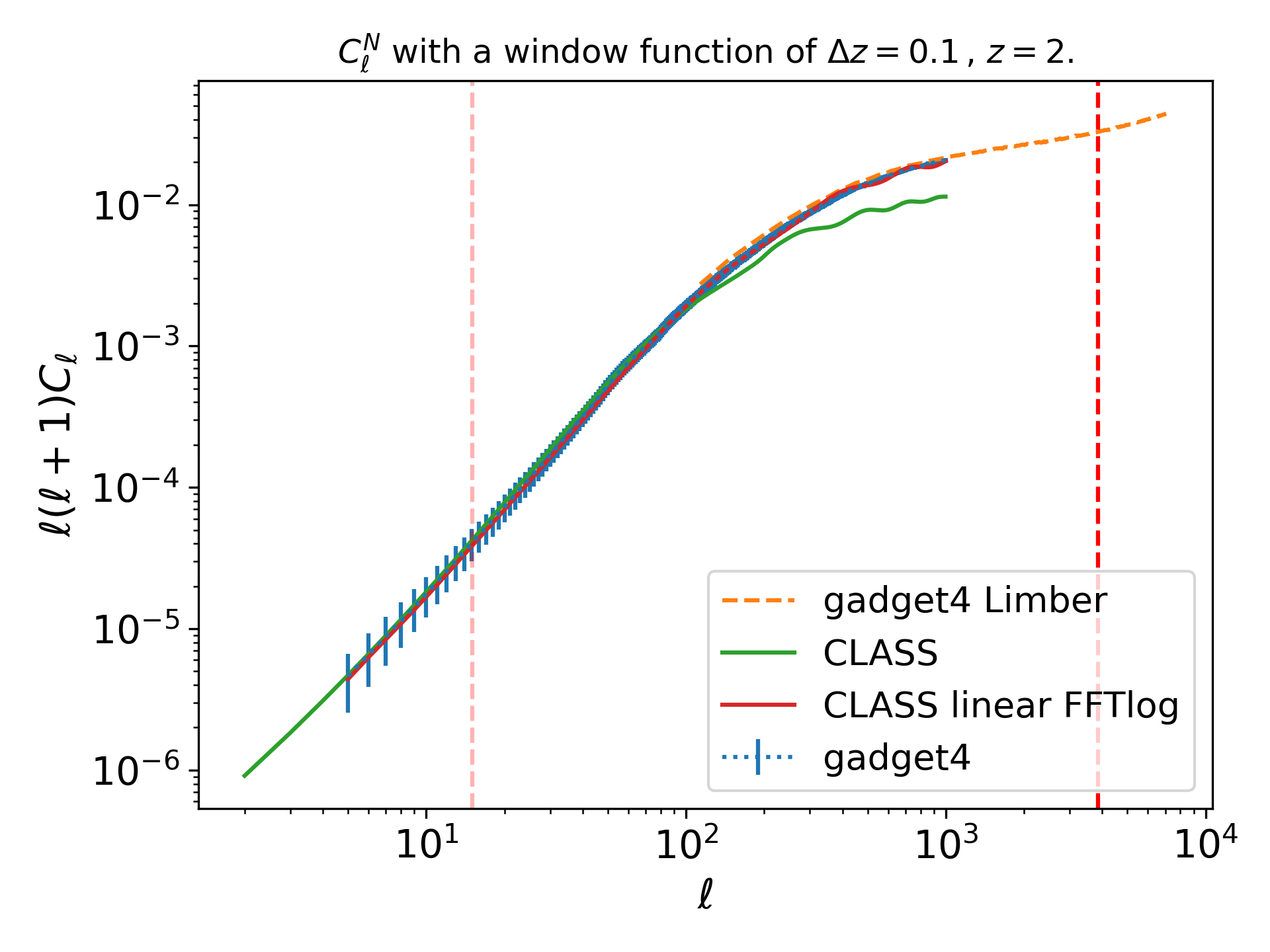}
    \caption{Angular power spectrum of matter for redshifts $z=0.5\,,1,\,2$. The power spectra are calculated in three different ways: using the \textsc{CLASSgal} code, using the \texttt{FFTlog} routine to obtain the angular spectra from the matter power spectrum of the simulation, and using the Limber approximation for the angular spectra from the simulation.
    The vertical red lines mark $2 \ell_\text{fund}$ and $\ell_\text{Nyq}/2$.}
    \label{fig:cl_validation}
\end{figure}

We compute the matter power spectra at redshifts $z=0.5,\,1\,,2$ with the \texttt{Pylians3} library%
\footnote{\href{https://github.com/franciscovillaescusa/Pylians3}{github.com/franciscovillaescusa/Pylians3}}~\citep[][]{2018ascl.soft11008V}, and test them against the  linear and nonlinear (\texttt{halofit}) matter power spectra derived from \texttt{class}.
The simulation snapshots are read in HDF5 format, and we employ a triangular-shaped cloud mass assignment scheme in \texttt{Pylians3} to generate the density fields.
We plot the results in figure \ref{fig:ps_validation}.
The curves obtained from the simulations match the expected behavior.
We subtracted the shot-noise contribution $P_N=V/N$ from the power spectrum measured from the simulation $\hat P$, that is, we show $ P = \hat P - V/N$. A derivation of this expression for Poissonian noise can be found in the references \cite{durrer_2020,Dodelson:2003ft}.

\begin{figure}
    \centering
    \includegraphics[width=0.5\linewidth]{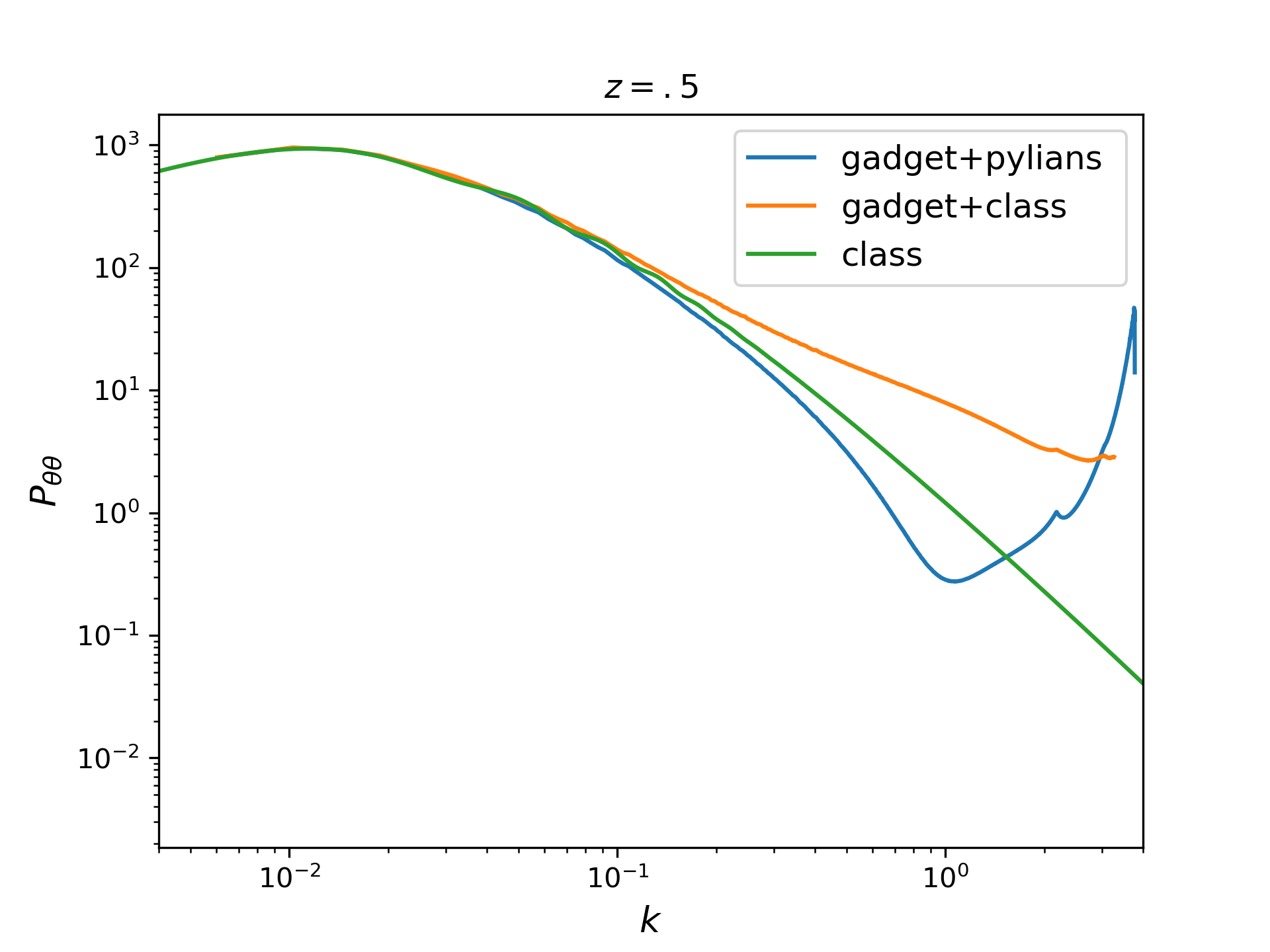}\hfil
    \includegraphics[width=0.5\linewidth]{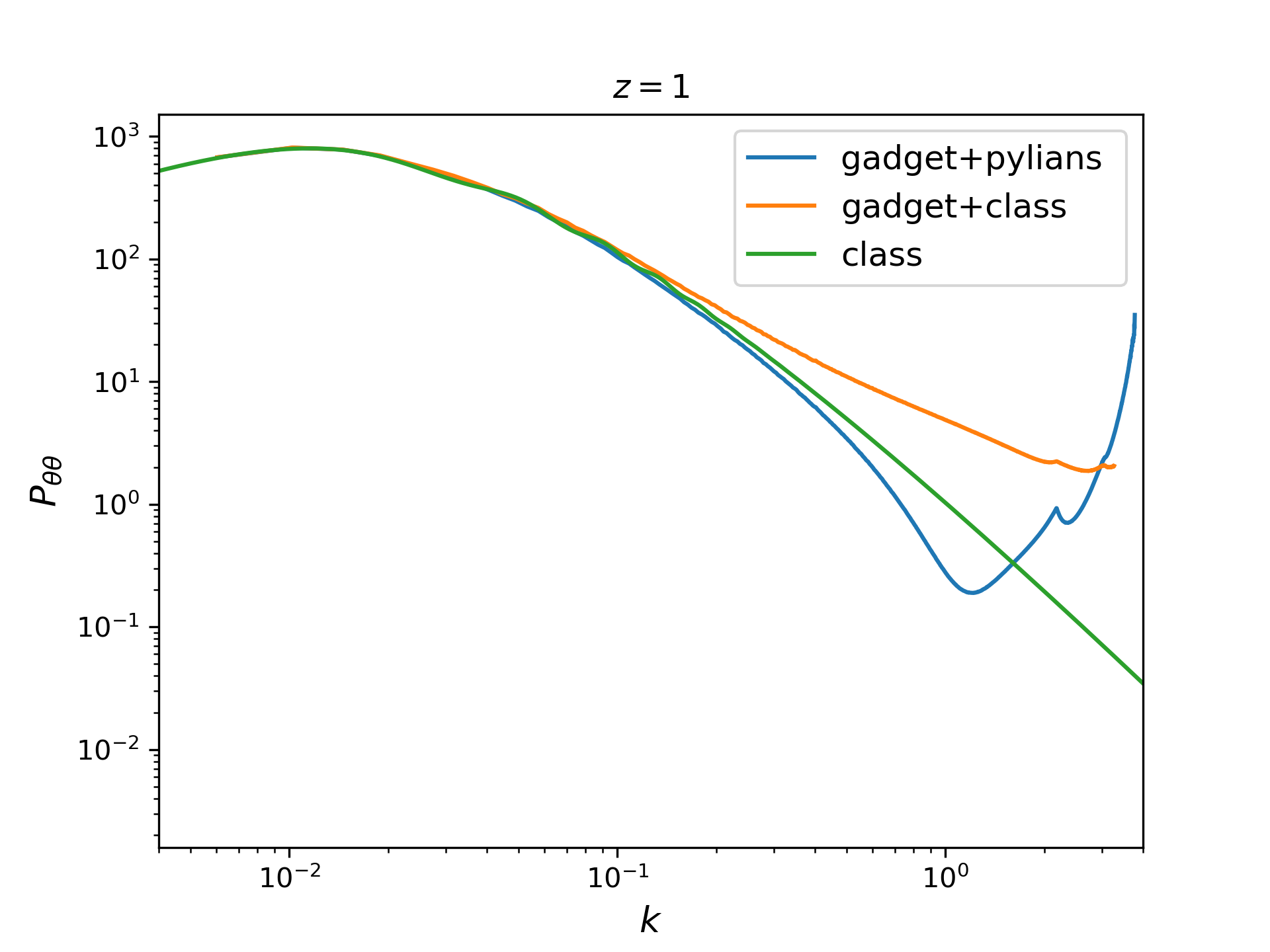}
    \includegraphics[width=0.5\linewidth]{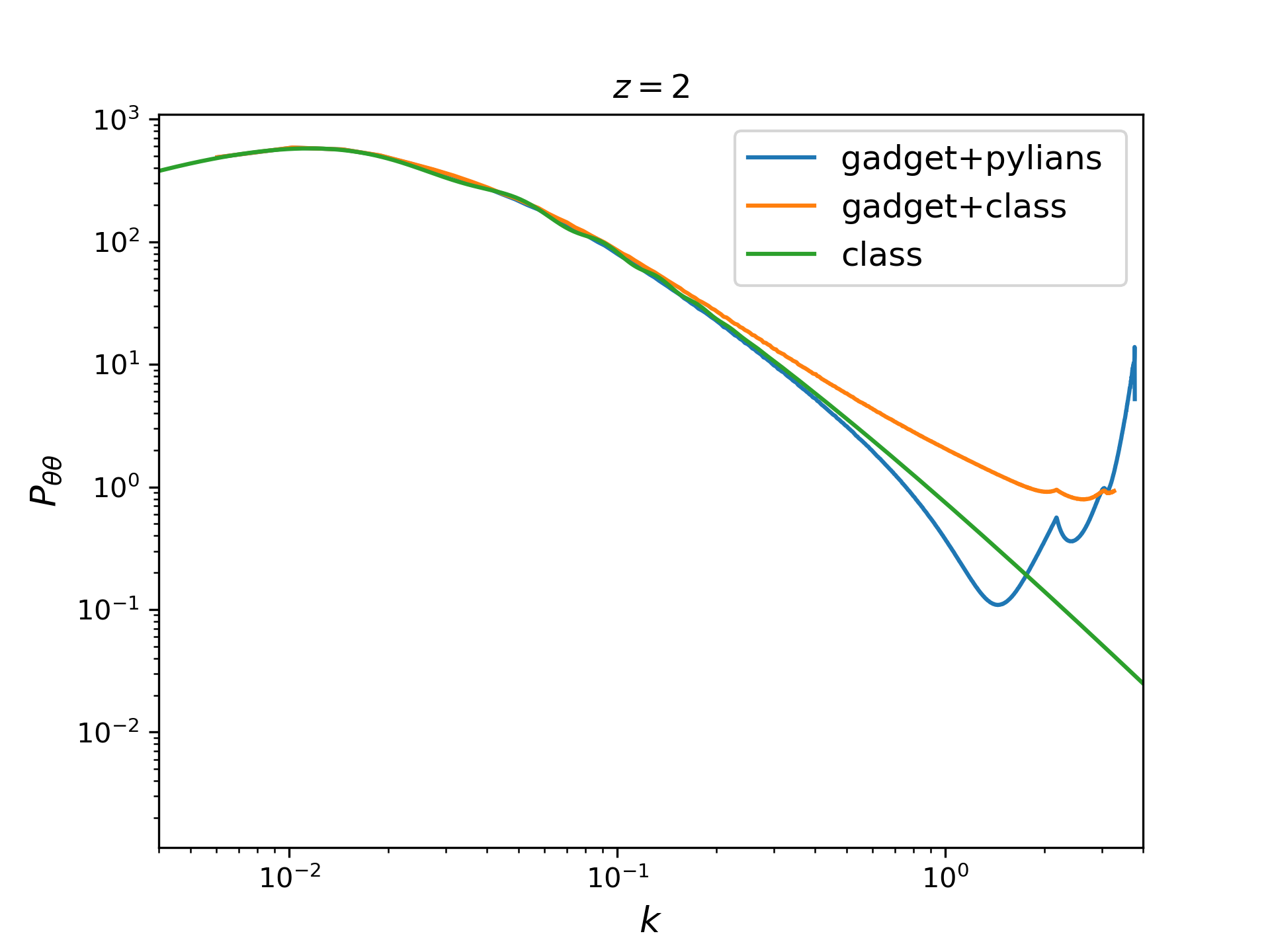}
    \caption{Velocity gradient power spectra $P_{\theta\theta}(k)$ obtained from the \texttt{class} code (red), from the \texttt{Pylians3} routine (blue) and from the continuity equation~\eqref{eq:continuity} (orange).}
    \label{fig:ps_theta_validation}
\end{figure}

To validate the routines used to compute the angular power spectra in equation~\eqref{eq:cl_drift}, we compare the results obtained from the simulation with the \texttt{FFTlog-and-beyond}  library
against those from the \texttt{class} code (the \texttt{classgal} module). %, ensuring the use of identical window functions and redshift slices.
Additionally, we include the Limber approximation as a consistency check for the angular power spectra derived from the simulations. The results of this comparison are presented in figure \ref{fig:cl_validation}.
The agreement is not perfect because \texttt{classgal} computes the angular power spectrum on the lightcone, while we are using the power spectrum of a given snapshot.
We checked the importance of non-linearities by feeding the linear power spectrum from \texttt{class} to the \texttt{FFTlog-and-beyond} routine: we found the expected suppression of power at smaller scales and lower redshifts.
We estimate the cosmic variance on the angular power spectrum via $\text{Var} \left[ C_\ell \right] = \frac{2 C^2_\ell}{2 \ell +1} $.

To avoid effects due to finite box size and resolution, we restrict our spectra to modes $k\in[2k_\text{fund},k_\text{Nyq}/2]$ and angular correlations to modes $\ell\in\left[2 \ell_{\rm fund} =\frac{2k_\text{fund}r(z)}{(1+z)},\ell_{\rm Nyq}/2 =\frac{k_\text{Nyq}r/2}{(1+z)}\right]$, above the fundamental frequency $k_\text{fun} = 2\pi/\text{Box Size}$ and below the Nyquist frequency $k_\text{Nyq} = N_\text{grid}\times k_\text{fun}$.

As an extra validation step on the calculation of divergence contributions to the final redshift drift spectra \eqref{eq:cl_drift}, we also calculate the velocity divergence $\theta$ power spectra for the three snapshots using the \texttt{Pylians3} code.
To assess the challenging convergence of $P_{\theta\theta}$~\cite{Esposito:2024qlo}, we evaluated it over a range of grid resolutions from 64 to 1024, finding that the results remain stable up to wavenumbers of 0.2–0.3.
The spectra agree with the \texttt{class} linear predictions up to wavenumbers 0.1--0.2, depending on the redshift, where it starts to diverge, as expected. The spectra are normalized so as to be easily comparable to results found in works on the velocity divergence spectra, such as the recent work on the non-linear regime of such spectra found in \cite{Esposito:2024qlo}.
We also compare the velocity divergence spectra obtained directly from \texttt{CLASS} and \texttt{gadget} to the velocity divergence spectra obtained by using the linear order continuity equation for the matter density \cite{Dodelson:2003ft} (also in Fig.~\ref{fig:ps_theta_validation})
\begin{equation}
    (aHf)^2 P_\delta(k) = P_\theta(k),
    \label{eq:continuity}
\end{equation}
where the $P_\delta$ spectrum is obtained in the same manner as the one found in figure \ref{fig:ps_validation}.We also find divergence in relation to the linear class prediction at $k\sim 0.1$--0.2.

\subsection{Redshift drift}

\begin{figure}
    \centering
    \includegraphics[width=0.5\linewidth]{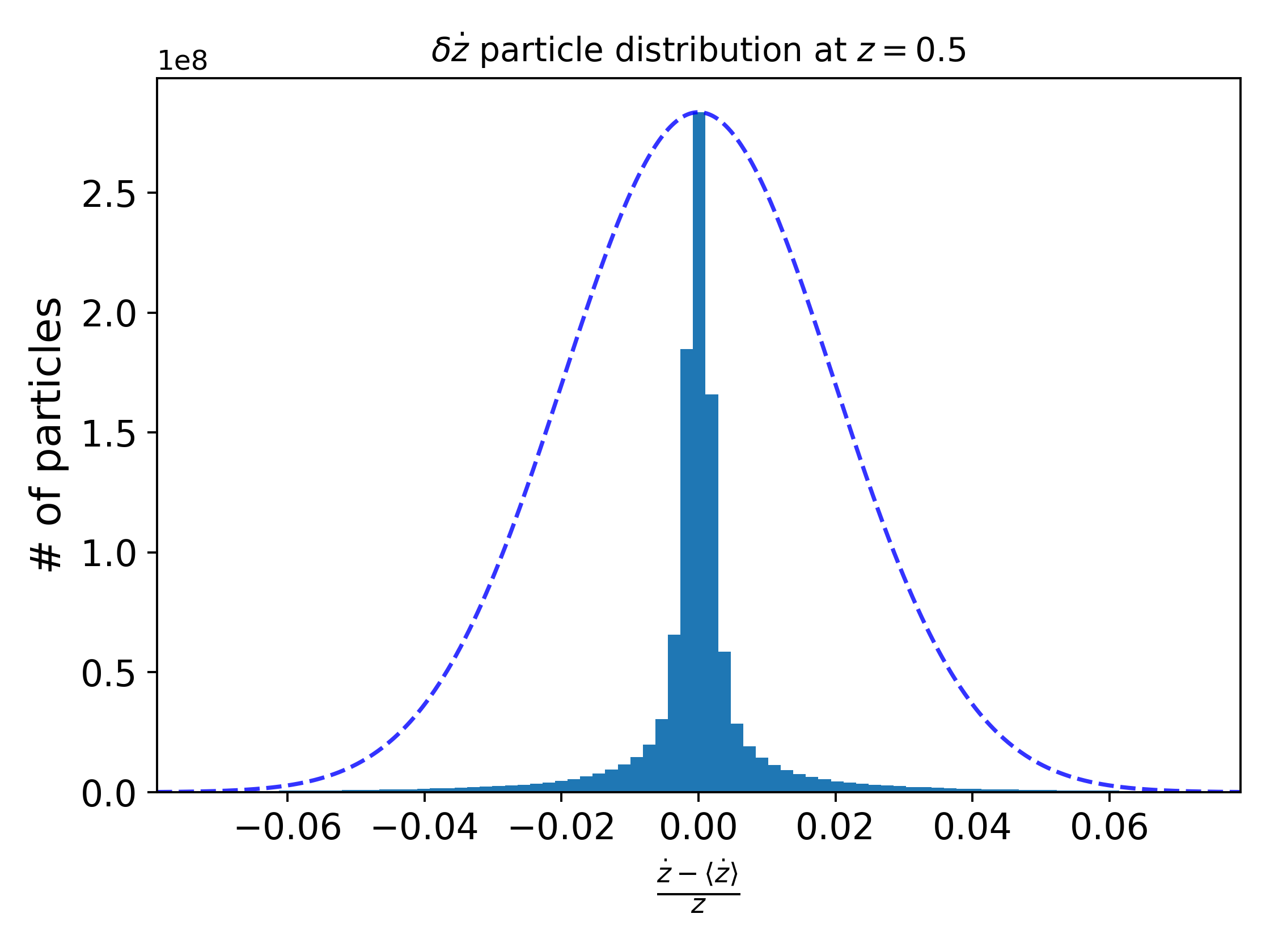}\hfil
    \includegraphics[width=0.5\linewidth]{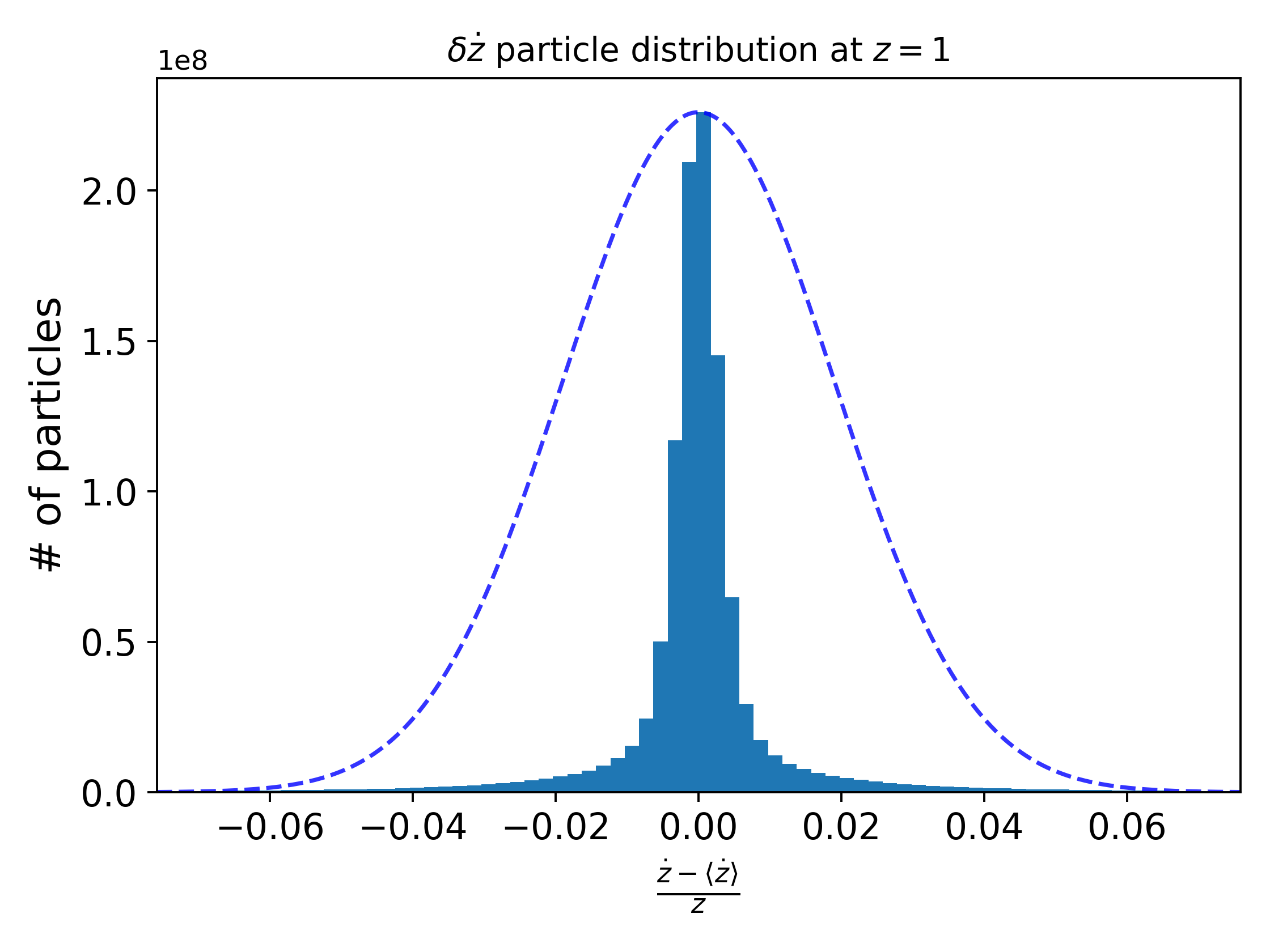}\par\medskip
    \includegraphics[width=0.5\linewidth]{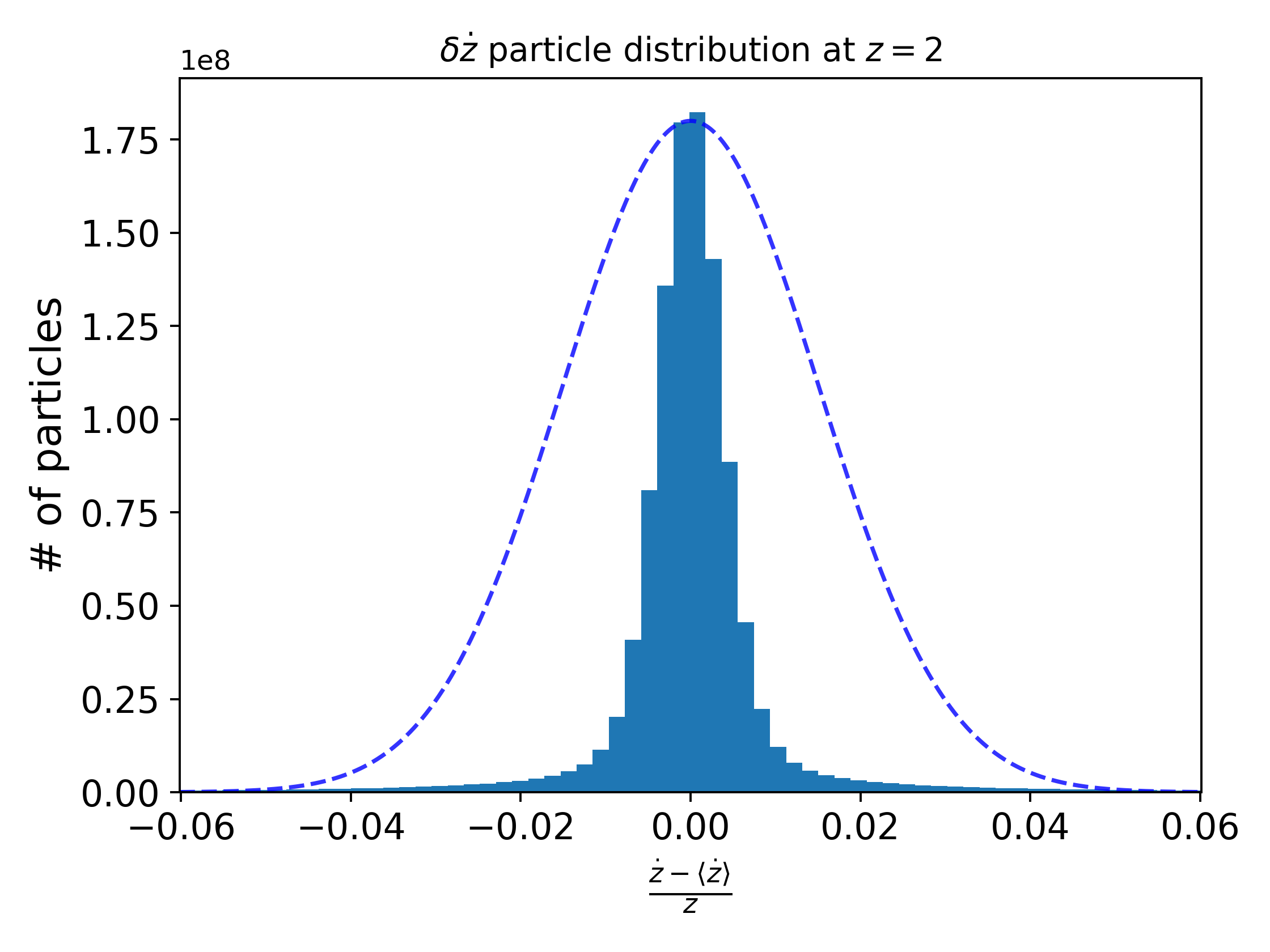}\hfil
    \caption{Histogram for the redshift drift fluctuations in the simulation box for redshifts $z=0.5,\,1,\,2$. We plot a normal distribution with the same standard deviation for comparison (dashed blue), highlighting a strong kurtosis.}
    \label{fig:drift_hist}
\end{figure}

To generate the 3D array of the redshift drift field, we discard potential terms and follow equation~\eqref{eq:drift_corr} to derive the drift expression from the acceleration and velocity fields.
Via \texttt{Pylians3} we obtain the $1024^3$ array with the total drift per voxel.
We then divide the total redshift drift field by the particle number field, in order to obtain the average redshift drift field.
It can happen that there are empty voxels; in this case we set the average redshift drift to zero.
The amount of empty voxels is less than $100$ in a total of $2^{30}$ voxels, such that the effect of this procedure is negligible.
%, which does not affect the final statistics.
We then subtract the average redshift drift in the snapshot from the redshift drift field and divide by the same average to obtain the redshift drift fluctuations of \eqref{eq:dz_fluct}. We plot the redshift drift fluctuation distribution histogram in figure \ref{fig:drift_hist} for the three redshifts analyzed.

\begin{figure}[ht]
\centering
\includegraphics[width=0.50\linewidth]{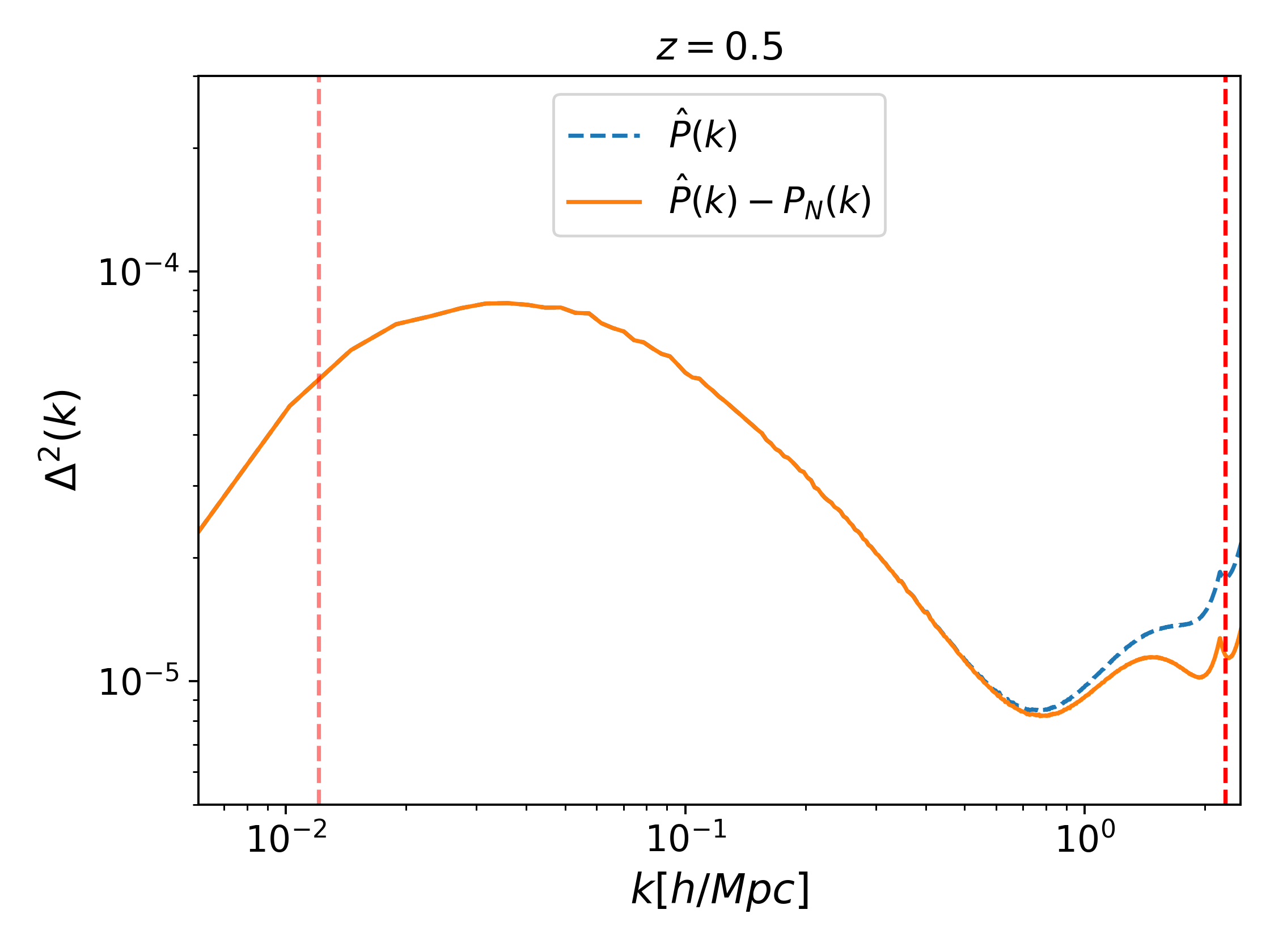}\hfil
\includegraphics[width=0.50\linewidth]{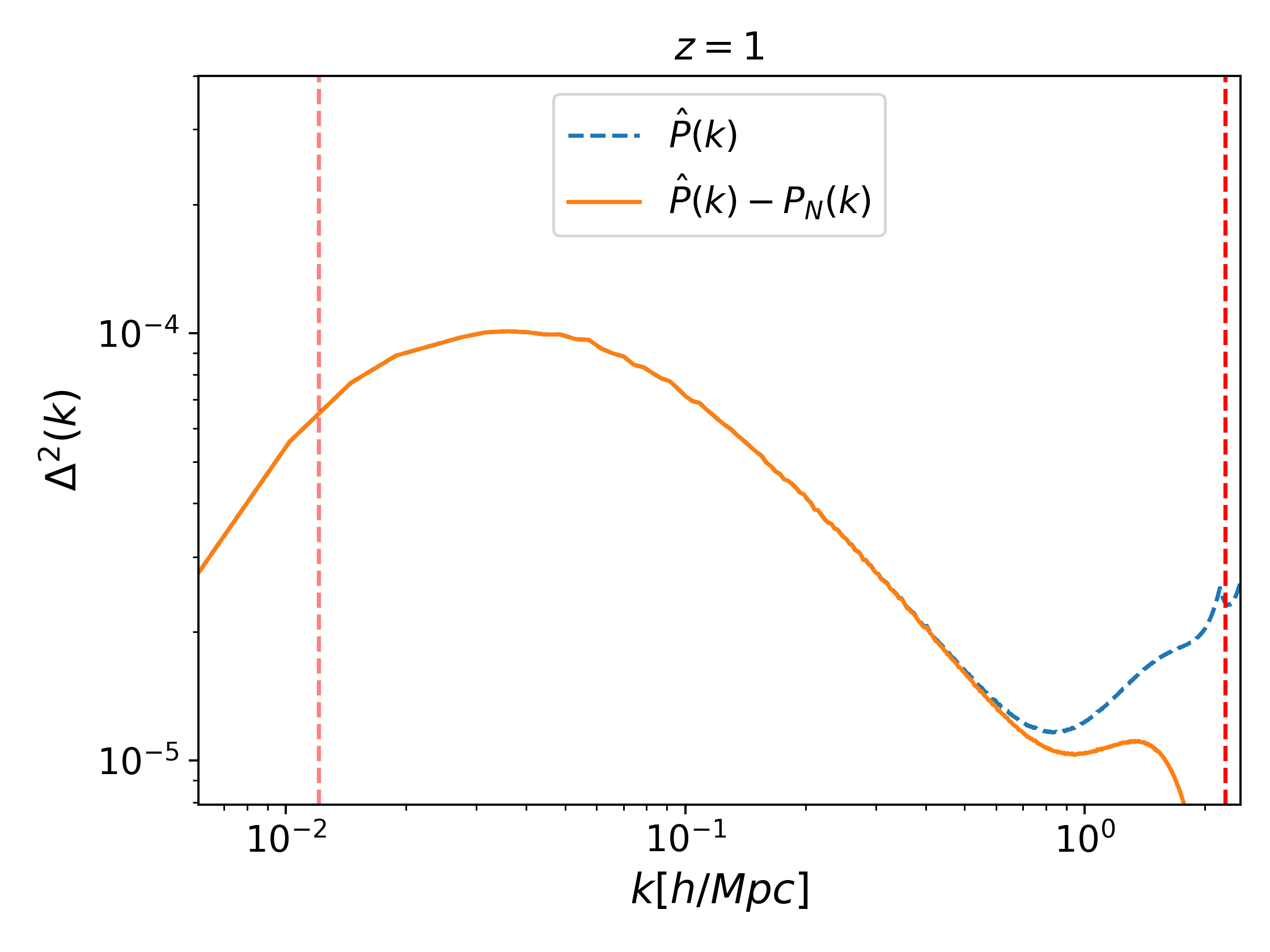}
\includegraphics[width=0.5\linewidth]{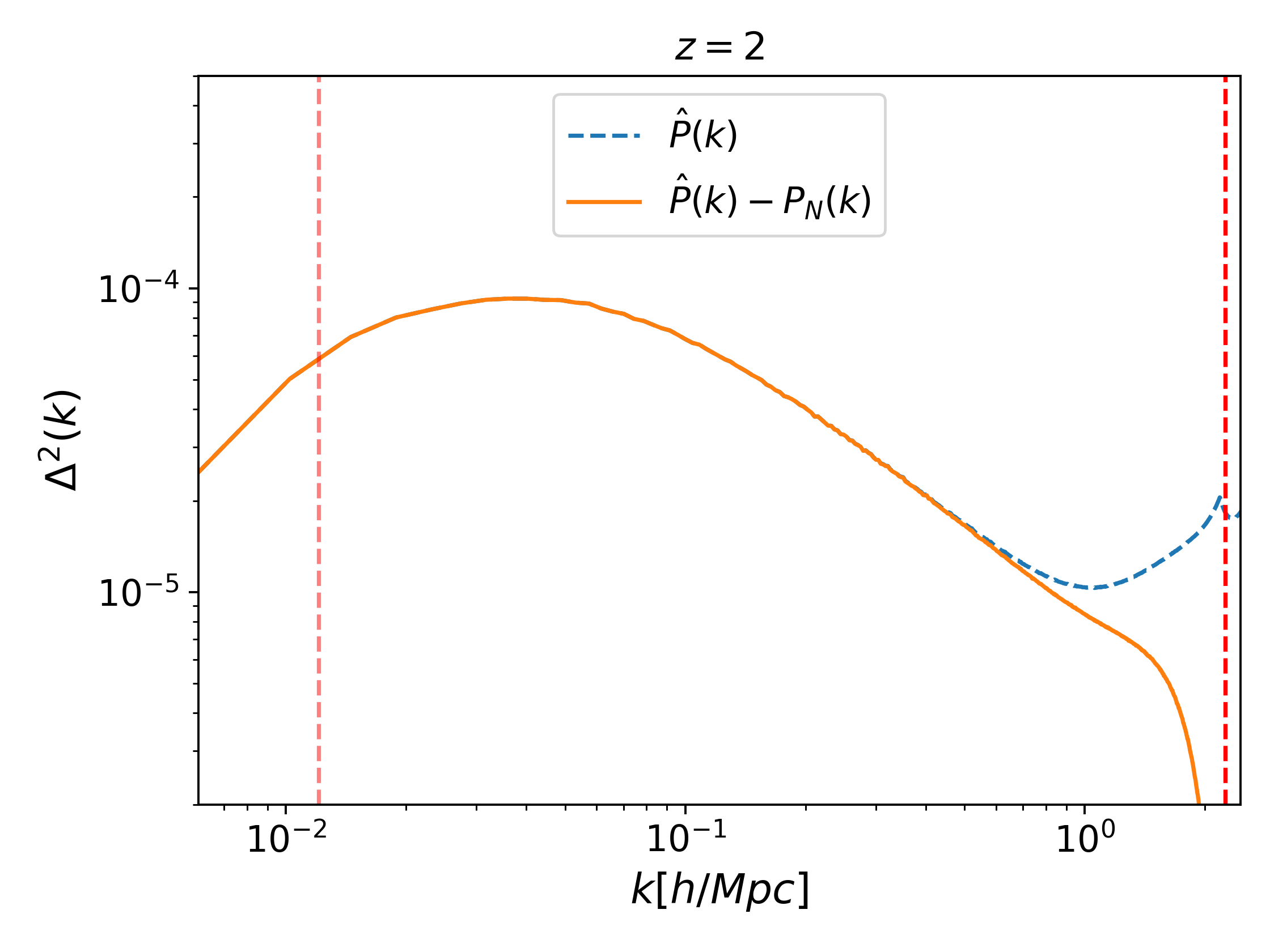}
\caption{Adimensional redshift drift power spectra at redshifts $z=0.5,\,1,\,2$ with and without shot noise. The solid lines give the estimated spectra minus the shot noise and the dashed lines give the full estimated spectra.
The vertical red lines mark $2k_\text{fund}$ and $k_\text{Nyq}/2$.}
\label{fig:ps_dz}
\end{figure}

The shot noise for the drift field can be computed following the formalism of \cite{Howlett:2019bky} such that
\begin{equation}
    \label{eq:sn_drift}
    P^{\delta \dot{z}}_{N} \simeq   \frac{V}{N} \langle\delta \dot{z}^2\rangle \,,
\end{equation}
that is, it is given by the matter power spectrum shot noise multiplied by the variance of the drift field. We then have
\begin{align}
P^{\delta \dot{z}} = \widehat{P^{\delta\dot{z}}}-P^{\delta \dot{z}}_{N} \,,
\end{align}
where we subtracted the shot-noise contribution from the power spectrum measured from the simulation $\hat P$.
We plot the dimensionless power spectra $\Delta^2(k,z)$ of the redshift drift in figure~\ref{fig:ps_dz} with and without subtraction of the shot noise. The subtraction effectively removes the excess power at small scales, $k \sim 2,h$/Mpc, at $z=2$, while at $z=0.5$ the correction is partially underestimated.

We see that the amplitude of the redshift drift fluctuations is $|\delta|\sim \sqrt{\Delta^2} \approx \mathcal{O}(10^{-2})$, and that the power spectrum is stronger at higher redshifts, where the cosmic velocity and acceleration fields in \eqref{eq:dz_fluct} are known to be larger, since velocity perturbations decay as a function of cosmic time in $\Lambda$CDM \cite{durrer_2020}. 
This result is in agreement with the results found in \cite{Bessa_2023} and around two orders of magnitude higher than the results found in \cite{koksbang_redshift_2023}, which also used a Newtonian simulation to obtain an estimate of the fluctuation amplitude. We note that the results in \cite{koksbang_redshift_2023} are based on an Einstein-de Sitter background cosmology, such that some of the discrepancies can be attributed to the effect of both the background cosmology and structure formation in $\Lambda$CDM, which is known to be slower due to the late time acceleration.

\begin{table}[h]
\begin{center}
\begin{tabular}{|l|l|l|}
\hline
Redshift& SN& SKA\\
\hline
$0.5$& $1.1\cdot 10^{-5}$&$3.2\cdot 10^{-4}$ \\
$1 $&$2.2\cdot 10^{-5} $&$1.8\cdot 10^{-3}$\\
$2 $& $3.8\cdot 10^{-5}$& $3.2\cdot 10^{-3}$ \\
\hline
\end{tabular}
\end{center}
\caption{Shot noise for the redshift drift power spectra and the expected noise for the redshift drift experiment with SKA1 and SKA2, taken from \cite{klockner2015realtimecosmology}. The value for $z=2$ is extrapolated from the expected noise at $z=1.5$.}
\label{tab:shot_noise}
\end{table}

For $N_\text{particles}=1024^3$ and the cosmic parameters given in table \ref{tab:parameters}, the values of the shot noise $P^{\delta \dot{z}}_{N}$ at the analyzed redshifts can be found in table \ref{tab:shot_noise}.
The expected shot noise in the SKA survey experiment aimed at measuring the redshift drift has been estimated in \cite{klockner2015realtimecosmology}, and by assuming that the survey will be able to precisely measure the fluctuations to constrain the background value of the redshift drift, we can compare the results of table \ref{tab:shot_noise} to the error due to shot noise predicted for the SKA1 and SKA2 experiments as found in \cite{klockner2015realtimecosmology,Bessa_2023}, which is expected to observe $N\sim10^7$ galaxies per redshift bin and the  noise term proportional to $1/\sqrt{N}$. Table \ref{tab:shot_noise} shows that the shot noise obtained from our work is small compared to the noise predicted from the survey capabilities, such that one does need to take into account the shot noise from the drift fluctuations when estimating the power spectrum.

\subsection{Comparison with \texttt{class} and discussion}
\label{sec:results}

\begin{figure}[t]
    \centering
    \includegraphics[width=0.5\linewidth]{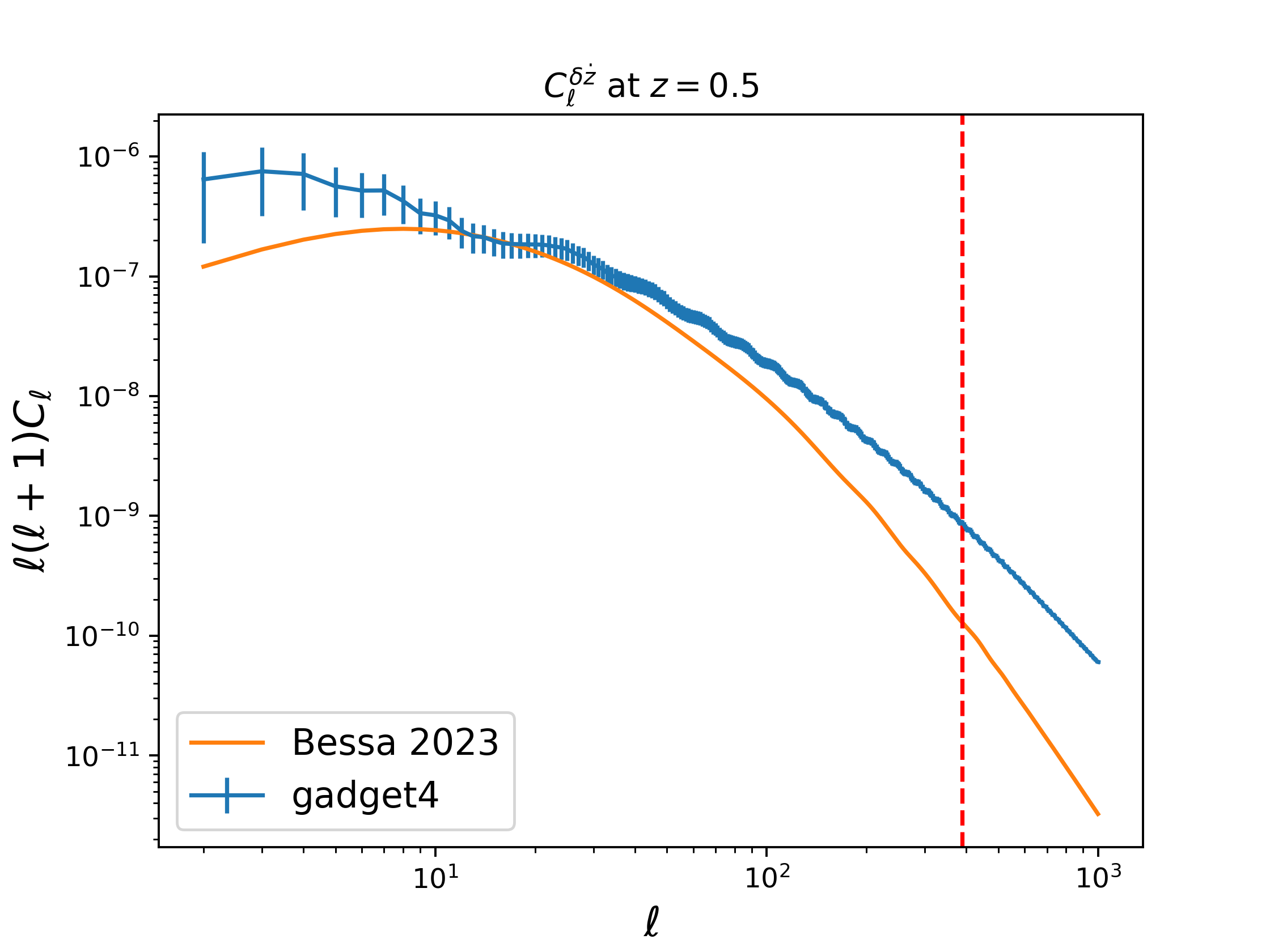}\hfil
    \includegraphics[width=0.5\linewidth]{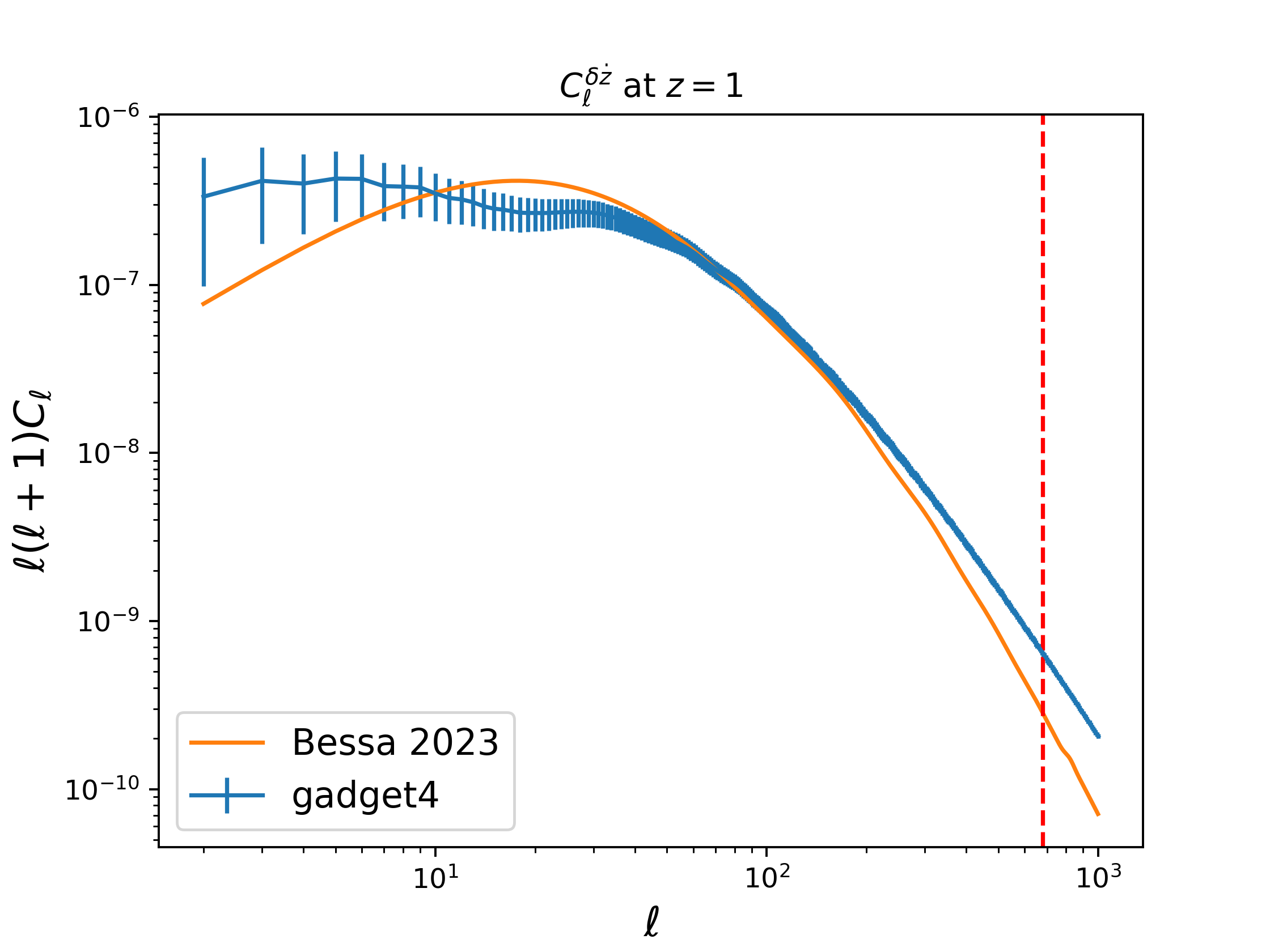}\par\medskip
    \includegraphics[width=0.5\linewidth]{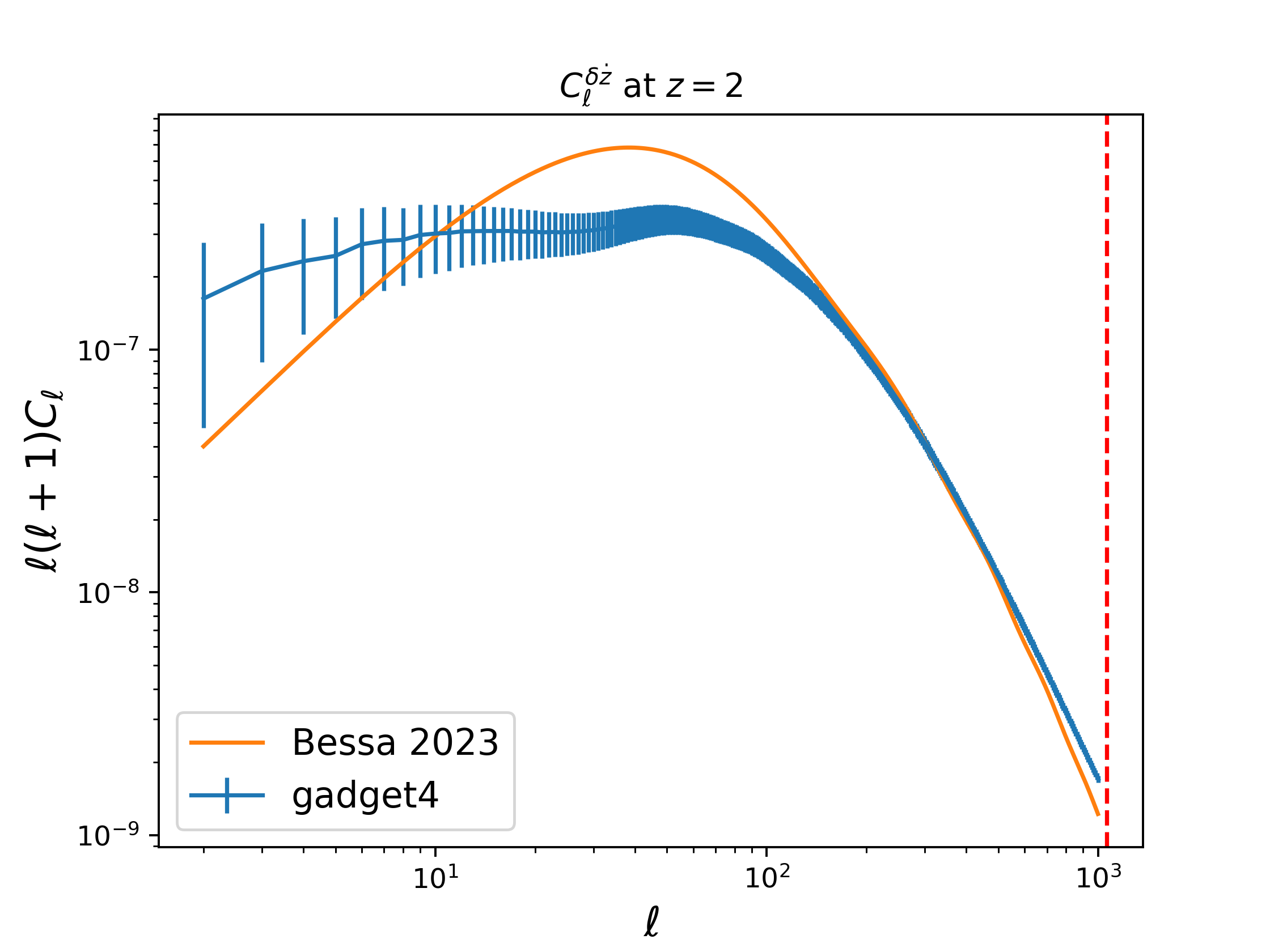}\hfil
    \caption{Redshift drift angular power spectra for redshifts $z=0.5,\,1,\,2\,$. The orange line represents the prediction for the angular power spectrum using the implementation in the \texttt{class} cosmology code. The blue line is relative to the \texttt{gadget4} snapshots. The vertical red slashed line marks the $\ell$ corresponding to the convergence scale obtained for the velocity spectra ($k=0.2$ Mpc/$h$).}
    \label{fig:cl_dz_gadget_class}
\end{figure}

The modified \texttt{class} code used to generate the angular power spectrum $C_\ell^{\delta\dot{z}}$ operates strictly within linear perturbation theory, meaning that nonlinear effects are not included in the redshift drift power spectra. However, since our simulations probe the nonlinear regime, deviations from linearity appear at smaller scales, leading to modifications in the slope of the angular spectrum compared to the linear spectra produced by \texttt{class}. We normalize the spectra derived from simulations, based on \eqref{eq:cl_drift}, to match the convention used in \cite{di_dio_classgal_2013}, enabling a proper comparison between the amplitude of the $N$-body angular spectra and the predictions from \texttt{class}. This normalization convention is detailed in Appendix \ref{sec:params} for reference.

In figure~\ref{fig:cl_dz_gadget_class}, we plot both the spectra obtained from the simulations and the spectra generated using the \texttt{class} implementation of the redshift drift fluctuations spectra.
Note that the power spectrum labeled `Bessa 2023’ employs the top-hat window function with $\Delta z = 0.1$ adopted in this work, rather than the Gaussian window function with $\Delta z = 0.01$ used in~\cite{Bessa_2023,Oestreicher:2025qcs}. Consequently, the resulting power spectrum differs from those shown in Fig.~3 of~\cite{Bessa_2023} and Fig.~11 of~\cite{Oestreicher:2025qcs}.

The spectra follow a similar behavior for different scales at different redshifts, with a particular good agreement for redshifts $z=1$ and $z=2$, where the perturbations are not deep into the nonlinear regime. We have checked that the window function \eqref{eq:window_func} size does not alter significantly the concordance between the two spectra, and also include the corresponding convergence scale for the velocity divergence power spectra obtained from the consistency checks in the previous section.
We also checked that the accuracy
in the numerical implementations does not alter the final result, such that numerical errors are under control in the routines.

In comparison with the \texttt{class} spectra, the simulated spectra differ by up to an order of magnitude at the largest scales ($\ell\lesssim10$) across all redshifts, with error bars generally not including the \texttt{class} results. At intermediary scales $10\lesssim\ell\lesssim100$ for redshifts $z=0.5$ and $z=1$ the spectra seem to be in good agreement, although there is a dip in the spectra obtained from \texttt{gadget} not found in the \texttt{class} linear spectra, changing the overall slope of the curve. This dip is more pronounced at redshift $z=2$, where the spectra differ within an order of magnitude. As expected the oscillations are more pronounced as the redshift gets smaller due to nonlinearities, such that for $z=0.5$ they are quite distinguishable in the spectra obtained from the simulations. At the smallest scales $\ell\gtrsim100$ for $z=1\,,2$ the spectra are in good agreement, particularly for $z=2$, where they seem to perfectly match up to the convergence scale $k=0.2$. For $z=1$ the agreement gets worse as $\ell$ increases, where it reaches a $20\%$ difference between the spectra. For $z=0.5$, into the non-linear regime, the disagreement between the spectra at the smallest scales is quite pronounced, where there is again at most an order of magnitude difference. Such scales at this redshift, however, are quite non-linear, and larger discrepancies are expected in relation to the linear \texttt{class} spectrum.
Although the plots point to a similar behavior and amplitude of the spectra, the discrepancy, in particular at small and large scales, needs to be further worked out and point to a need for a more accurate modeling of the spectra, such as including light-cone integration and ray-tracing.
Although the plots indicate similar overall behavior and amplitude, the discrepancies at small and large scales, as well as the fact that at $z=2$ the small scales agree better than the large linear ones, require further investigation. These differences point to the need for a more accurate modeling of the spectra, including light-cone integration and ray-tracing.
We also list here some caveats and differences between our results and the ones previously obtained in the literature that could contribute to the discrepancy found in the power spectra:
\begin{itemize}
    \item The calculation of clustering is based on the simulation snapshot drift power spectrum rather than on integration over the light-cone, as done by the \texttt{class} code. This could explain part of the difference between the two spectra, as the implementation of the methods is quite different.
    \item The \texttt{class} code does not directly calculate the acceleration power spectra \cite{DiDio:2013bqa,Bessa_2023}, instead relying on the first order conservation equation, whereas the \texttt{gadget4} code calculates it directly from the gradient of the potential fields \cite{Springel:2020plp}, which deviates the most at the smallest scales, where non-linearities enter the spectra.
    
    \item The methods used in \cite{koksbang_redshift_2023,Koksbang:2024xfr} to obtain the amplitude and power spectra of the fluctuations involve ray-tracing of the particles light-cones, and the integrals found in~\eqref{eq:cl_drift} are done along the light-cone in the \texttt{class} code, such that finite Box Size effects due to the simulation limitations could impact the overall shape of the spectra at the larger scales.

    \item It is possible that the mapping found in \cite{green_newtonian_2012} between Relativistic and Newtonian Cosmologies, used throughout the text to map the relativistic equations to the $N$-body quantities, breaks down when dealing with quantities at the order of magnitude of the redshift drift fluctuations.
\end{itemize}
These caveats, although not extensive, provide a path forward to further investigate the discrepancies between the works. In particular, the fact that the acceleration field cannot be directly validated through \texttt{class} needs to be addressed in order to properly check that the angular spectra being calculated is agreeing within the expected numerical error. This effort entails modifications in the \texttt{class} source code, which we leave for future work.

%The order of magnitude agreement between our results and the results found in \cite{Bessa_2023} also support the approximations done in equations~\eqref{eq:dz_fluct} and \eqref{eq:cl_drift} to arrive at the final expression for the spectra. The potential terms, which are included in the analysis done in \cite{Bessa_2023}, can be safely neglected in relation to the dominant acceleration and velocity terms, at all scales.

In our simulation setting, we were able to estimate the amplitude and scaling of the power spectrum of the drift fluctuations through observations of the cosmic velocity and acceleration fields, which together with the galaxy number count spectra, as discussed in~\cite{Bessa_2023}, could help in detecting a statistically significant signal of the drift by constraining contaminating effects due to the peculiar motion of sources.

\section{Conclusions}
\label{sec:conc}

In this paper, we derive the redshift drift power spectrum at different redshifts from an $N$-body simulation using the \texttt{Gadget4} code. To obtain the power spectrum from the simulation, we model it using the theoretical prediction found in \cite{Bessa_2023} and numerically calculate it using the velocity and acceleration fields from the \texttt{Gadget4} snapshots, and power spectrum and FFT routines found in the \texttt{pylians} and \texttt{FFTlog-and-beyond} python libraries. We compare and validate our code and simulation using the Einstein-Boltzmann solver \texttt{class}, and once validated, compare our results to the numerical predictions obtained using the same code in \cite{Bessa_2023}. Our work provides a baseline methodology for future modeling of the redshift drift using $N$-body simulations and derivations of one and two point statistics from simulation data, while needing further improvements in accurately modeling the spectra.

The redshift drift spectra we obtain provide a first step in accurately modeling the redshift drift spectra using $N$-body simulations. When comparing our results to the \texttt{class} code and previous studies discrepancies appear in the shape at large scales and low redshifts, which demand further development of our methods. We discuss possible causes, including the treatment of light-cone quantities, the assumed background cosmology, and the computation of particle acceleration fields.
Further work with consistent background cosmologies, ideally carried out directly on the past light-cone, is required to enable robust comparisons across numerical simulations of the effect. Nevertheless, our study represents a first step toward modeling drift fluctuations with data from future surveys. A more complete assessment of differences in power spectrum estimates between methods will require systematic analyses, including fully relativistic simulations such as those in \cite{Koksbang:2024xfr}.
Our work agrees with previous literature in that the fluctuations of the drift can impact the background measurement, as found in \cite{Bessa_2023,uzan_time_2008}, in particular in measuring the velocity and acceleration fields distribution over the sky. Although of order $10^{-2}$ on the background drift, which is already small, it is within the capabilities of surveys such as SKA~\cite{Bessa_2023,Koksbang:2024xfr}.

Finally, our simulation assumes an idealized dark matter plus $\Lambda$ only Universe, with structure formation and peculiar motion of sources with no baryonic effects or feedback, and a simulation box size of $1\textrm{Gpc}$. The SKA radio telescope will have a sky coverage of $75\%$ and probe redshifts $z\in [0,5]$ in the radio, HI and 21cm regions, observing baryonic feedback in regimes previously unseen, with measurements of the drift up to redshift $z\approx 1.5$~\cite{Marques:2023qqu}. Larger simulations, including baryonic and hydrodynamic effects in structure formation and the momentum fields, are to be done to prepare for the upcoming observational challenge in measuring the drift with next-generation surveys. 
Such a task is beyond the current computational capabilities of the current work, and we leave it for future research.

\section*{Acknowledgements}
We are very thankful for the insights and commentaries by Ruth Durrer and Matteo Esposito.
It is a pleasure to acknowledge, Enea Di Dio, Cullan Howlett, Francisco Villaescusa-Navarro and Volker Springel for their valuable discussions.
PB acknowledges financial support from Coordenação de Aperfeiçoamento de Pessoal de Nível Superior (CAPES) and
Fundação de Apoio à Pesquisa do Espírito Santo (FAPES) for his PhD and visiting PhD fellowship.
VM thanks CNPq (Brazil) and FAPES (Brazil) for partial financial support. TC is supported by the Agenzia Spaziale Italiana (ASI) under - Euclid-FASE D  Attivita' scientifica per la missione - Accordo attuativo ASI-INAF n. 2018-23-HH.0, by the PRIN 2022 PNRR project "Space-based cosmology with Euclid: the role of High-Performance Computing" (code no. P202259YAF), by the INFN INDARK PD51 grant, and by the FARE MIUR grant `ClustersXEuclid' R165SBKTMA.
We acknowledge the computing centre  of Cineca and the use of the Santos Dumont supercomputer of the National Laboratory of Scientific Computing (LNCC, Brazil). The authors would like to acknowledge the use of the computational resources provided by the \href{https://computacaocientifica.ufes.br/scicom}{Sci-Com Lab} of the Department of Physics at UFES, which was funded by FAPES, CAPES and CNPq.

\appendix

\section{Limber approximation}
\label{sec:limber}

The Limber approximation makes use of the fact that the spherical Bessel functions $j_\ell(x)$ peak around $x = \left[\ell(\ell+1)\right]^{1/2}\approx \ell+ 1/2$  \cite{LoVerde:2008re}. For large $\ell$, this peak is even more pronounced, such that one may take $j_\ell(x)\xrightarrow{\ell\gg1} \sqrt{\frac{\pi}{(\ell+1/2)}}\delta(x-\ell-1/2)$, such that, for a function $f(k)= \int dx'f(x')j_\ell(kx')$, we have
\begin{equation}
    \label{eq:limber}
    \int_0^\infty dk f(k) = \int_0^\infty\int dkdx' f(x')j_\ell(kx') \approx \int_0^\infty dk f\left(\ell+\frac{1}{2} \right).
\end{equation}
This approximation has been used extensively on scales $\ell\gtrsim \mathcal{O}(10)$ for cosmological angular correlations, including in \texttt{class} \cite{di_dio_classgal_2013}.

\iffalse
\section{Flat sky approximation}
\label{sec:flat}

In the flat-sky approximation we assume that we are observing a small enough area of the sky such that the curvature of the celestial sphere can be neglected. More precisely, we substitute the spherical coordinates over the whole celestial sphere by
%
\begin{equation}
\label{eq:spher_coord}
    \bold{n} = r(z)(\theta,\phi) \longrightarrow \bold{e_z} = r(z)(x,y),
\end{equation}
where $\bold{e_z}$ is the unitary vector in a given coordinate in the simulation box, assumed to be the observer or survey reference direction, and $(x,y)$ are 2D coordinates in the plane perpendicular to $\bold{e_z}$. For small scales, this approximation is very accurate, as smaller areas of the sky are less affected by curvature effects. In this approximation, the modes $\ell$ are conjugate to the coordinates $\alpha = (x,y)$, such that they are related by a 2D Fourier transform. See \cite{Matthewson:2020rdt,Jelic-Cizmek:2020jsn} for a comprehensive discussion of the flat-sky approximation.

%this is for super think shells
%the authors show that the accuracy for self correlations and modes $\ell>2$ at a given redshift is within $0.3\%$ of the numerical accuracy of the \texttt{class} code for galaxy number counts, well beyond the needed accuracy for the derived power spectra in this paper.
\fi

\section{Units}
\label{sec:params}
%In this appendix we provide a compendium of the units and their conversion factors involved in the calculation of the angular Power Spectrum from the simulation snapshots and other numerical routines in the paper.

As the \texttt{Gadget4} code works with internal code units, one needs to convert from internal units to physical units and then map these fields to the physical fields found in equation \eqref{eq:cl_drift} through the conversion factors. We list the conversion factors and units used throughout the paper in the following table, and refer the reader to the \texttt{Gadget4} documentation at \href{https://wwwmpa.mpa-garching.mpg.de/gadget4}{wwwmpa.mpa-garching.mpg.de/gadget4} for a comprehensive list of the conversion factors.

\begin{table}[h]
\label{tab:units}
    \centering
    \setlength{\tabcolsep}{7pt}
\renewcommand{\arraystretch}{1.3}
    \begin{tabular}{|c|c|c|}
    \hline
   Simulation field [\texttt{Gadget4} units] & Physical field [physical units] & Conversion factor  \\
    \hline
    $h\,[10^{-2}\,{\rm km/s/Mpc}]$ & $H\,[{\rm km/s/Mpc}]$ & $10^{-2}$ \\
    $\bold{v}\, [{\rm cm/s}]$ & $\bold{v}\,[{\rm km/s}]$  & $10^{-5}\cdot a^{1/2}$ \\
    $\bold{a}\, [{\rm km^2/s^2/(Mpc/h)}]$& $\dot{\bold{v}}+ H\bold{v}\, [{\rm km/s^2/Mpc}]$ & $ h\cdot(\rm {km/s)^{-1}}$\\
    \hline
    \end{tabular}
    \caption{ Fields given in simulation units and physical units, and conversion factor from simulation to physical units. }
    \label{tab:my_label}
\end{table}

Furthermore, the \texttt{class} code outputs the angular power spectra of fields $X$ using the convention
\begin{equation}
    \label{eq:class_spectra} 
    \frac{\ell(\ell+1)}{2\pi}C_\ell^{XX}= \left(\frac{\ell(\ell+1)}{2\pi}\right) 4\pi\int \frac{dk}{k}\Delta^2_\ell(k,z),
\end{equation}
where $\Delta^2(k,z)$ is given by the $|F_\ell|^2$ in expression \eqref{eq:cl_theo}.

\bibliographystyle{JHEP}
\bibliography{biblio}

\end{document}